\documentclass{aa}  

\usepackage{graphicx}
\usepackage{txfonts}
\usepackage{ulem}
\usepackage{amsmath}
\usepackage{booktabs}
\usepackage{lastpage}

\usepackage{natbib}
\usepackage[colorlinks=true,linkcolor=blue, citecolor=blue]{hyperref}

\usepackage{tipa}

\newcommand{\Href}{h_\mathrm{ref}}
\newcommand{\rref}{r_\mathrm{ref}}

\begin{document}

   \title{DRAGyS - A comprehensive tool to extract scattering phase functions in protoplanetary disks}

   \subtitle{Disk ring adjusted geometry yields scattering phase function}
   \author{M. Roumesy
          \inst{1}
          \and
          F. Ménard\inst{1}
          \and
          R. Tazaki\inst{1, 2}
          \and
          G. Duchêne\inst{1, 3}
          \and
          L. Martinien\inst{1}
          \and
          R. Zerna\inst{1}
          }

   \institute{Univ. Grenoble Alpes, CNRS, IPAG, 38000 Grenoble, France
   \and  Department of Earth Science and Astronomy, The University of Tokyo, Tokyo 153-8902, Japan
   \and Astronomy Department, University of California Berkeley, Berkeley CA 94720-3411, USA}
   \date{Received --; accepted --}

  \abstract
  {The early stages of planet formation, involving dust grain growth and planetesimals formation, remain shrouded in mystery. The analysis of the Scattering Phase Function (SPF) measured in disks surrounding young stars holds great potential for revealing crucial information about dust grain properties. Given the rapidly increasing number of high-quality datasets available, an efficient method to extract the SPF is required.}
  {We developed {\tt{DRAGyS}} (Disk Ring Adjusted Geometry yields Scattering phase function),
  a tool designed for the quick and comprehensive analysis of protoplanetary disks in which gaps and rings are present. {\tt{DRAGyS}} directly estimates the disk geometry and extracts the total and polarized SPF from scattered light images, without requiring any radiative transfer modeling, a limitation of previous efforts.}
  {Key disk parameters — inclination, position angle, aspect ratio — are obtained by fitting ellipses to the disk intensity peaks from the ring surface, assuming the disks are circular. We validated the method using simulated disk images and then applied it to archival polarized-intensity images of nine images for six protoplanetary disks. {\tt{DRAGyS}} also provides a method to correct for the effect of limb brightening on the SPF.} 
  {{\tt{DRAGyS}} recovers well the injected geometry and the SPF from synthetic images where the parameters are known. When compared to previously published results extracted from images without taking into account limb brightening, {\tt{DRAGyS}} yields similar results for the inclination, disk position angle, and SPF. We show that the effect of limb brightening on the SPF is significant, with consequences for the inference of dust properties.}
  {{\tt{DRAGyS}} takes advantage of a fast and purely geometrical approach to estimate ringed-disk geometries. It allows the efficient extraction of SPF either globally or by sectors, allowing it to deal with disk asymmetries. By bypassing the need for a full modeling of the disk geometry before SPF extraction,  {\tt{DRAGyS}} is well suited to study large samples of disk images.}

   \maketitle

\begin{figure*}[htbp]
    \centering
    \includegraphics[width=\textwidth]{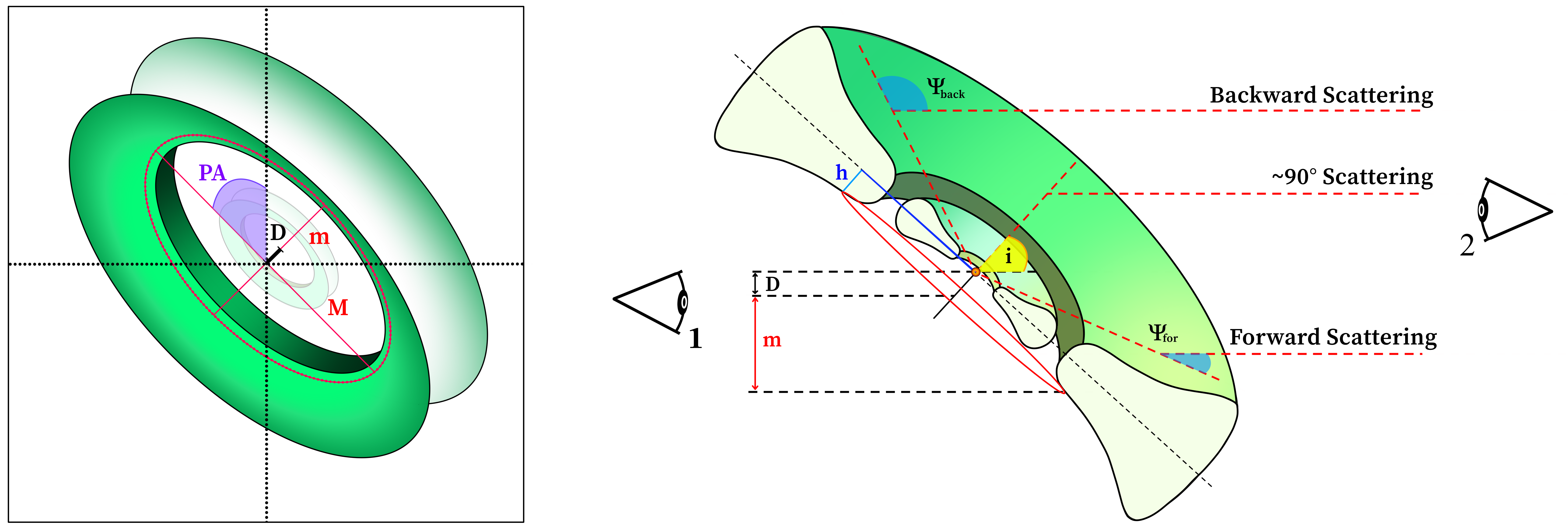}
    \caption{Illustration of the {\tt{DRAGyS}} method, illustrating the ellipse estimation of the disk (red), and the geometrical method to recover the different structure parameters.
    Left: image as seen by the observer, with the ellipse in red and different estimated quantities such as semi-minor and -major axis $m$ and $M$, the position angle $PA$ and the projected distance between star and major axis $D$ used to compute the aspect ratio. 
    Right: A sketch of protoplanetary disk observation, the point of view of observer 1 highlights the geometrical process to recover the disk geometry, namely the scattering surface height $h$ at a radius $r_{\mathrm{mid}}$ and the inclination $i$. The point of view of observer 2 illustrates the definition of the scattering angle.}
    \label{DiskScheme}
\end{figure*}
\section{Introduction}
Recent observations confirm that protoplanetary disks are the birthplaces of planets \citep{keppler_discovery_2018,pinte_kinematic_2018,teague_kinematical_2018}. Beyond these fascinating discoveries, instrumental improvements have led to the detection of a wide morphological diversity in disks, with spiral, gaps, and rings, being prominent and reflecting the complex planetary formation process \citep{benisty_optical_2022}. Indeed, submillimeter-sized dust grains must agglomerate to form planetesimals in a very short time to explain the birth of planets \citep{drazkowska_planet_2022}. The size and structure of dust aggregates influence their ability to grow \citep{johansen_multifaceted_2014}, and understanding these properties is important to better track the evolution of dust in disks.

The interaction of an electromagnetic wave with a dust particle depends on the wavelength, on the size and shape of the particle, and on the refractive index of the material. Therefore, measuring the Scattering Phase Function (SPF), describing the intensity of scattered light as a function of the scattering angle \citep[e.g.,][]{bohren_absorption_1983}, can be a powerful tool to study the dust structural properties \citep[e.g.][]{ginski_direct_2016, stolker_scattered_2016, tobon_valencia_scattering_2022, ginski_observed_2023, tazaki_fractal_2023, tobon_valencia_scattering_2024, chen_multiband_2024}.

In optically thin debris disks, where only single scattering is relevant, the interpretation of extracted SPFs is direct \citep{milli_near-infrared_2017}. The potential for significant absorption as well as multiple scattering and geometric complexity in protoplanetary disks makes the interpretation of their SPF more challenging. Yet, the image quality (angular resolution, contrast) provided by modern adaptive optics enhanced imaging instruments has led to a dramatic increase in the number of star and disk systems where SPFs can be reliably extracted \citep{ginski_observed_2023}. This opens the way to a deeper exploration of dust grain properties in the early stages of their evolution.

To date, few tools have been developed to extract the SPF from protoplanetary disk surfaces, such as {\tt{Diskmap}} \citep{stolker_scattered_2016}. However, these tools require a prior knowledge of the disk geometry, which is generally obtained by a customized ring fitting \citep{boer_multiple_2016, avenhaus_disks_2018}, a process we will generalize here, or by full radiative transfer model fitting. In the latter, the time-consuming model-fitting process is an obstacle to large-scale studies of dust grains in protoplanetary disks using SPF. To overcome this constraint, we developed a new comprehensive tool, {\tt{DRAGyS}}, to analyze the structure and scattered light properties of protoplanetary disks without prior information. Our approach stands out by its ability to estimate the disk geometry directly from an image, and directly extract the SPF. This method offers a fast and efficient alternative to existing methods, making it particularly well-suited to the study of large samples of protoplanetary disks, and a better understanding of the properties of dust grains they contain.

In Section \ref{S2_GeoFit}, we present the method to estimate the geometrical parameters of ring-shaped disks, i.e., their inclination, position angle, and aspect ratio. Then, we outline the steps involved in the extraction of the SPF from the surface of the protoplanetary disks. In section \ref{S3_SimulationPart}, we first apply our new tool to synthetic images of disks, then evaluate the impact of limb brightening and show how it can be corrected for. In section \ref{S4_RealData}, we test {\tt{DRAGyS}} on a sample of 6 ring-shaped disks observed in the near-infrared. These disks are chosen because their SPF have already been extracted and therefore serve as useful benchmarks \citep{ginski_observed_2023}. In section \ref{S5_Discussion}, we describe the advantages and limitations of this new tool. Conclusions are summarized in section \ref{S6_Conclusion}.

\section{Methodology} \label{S2_GeoFit}
\begin{figure*}
    \centering
    \includegraphics[width=\textwidth]{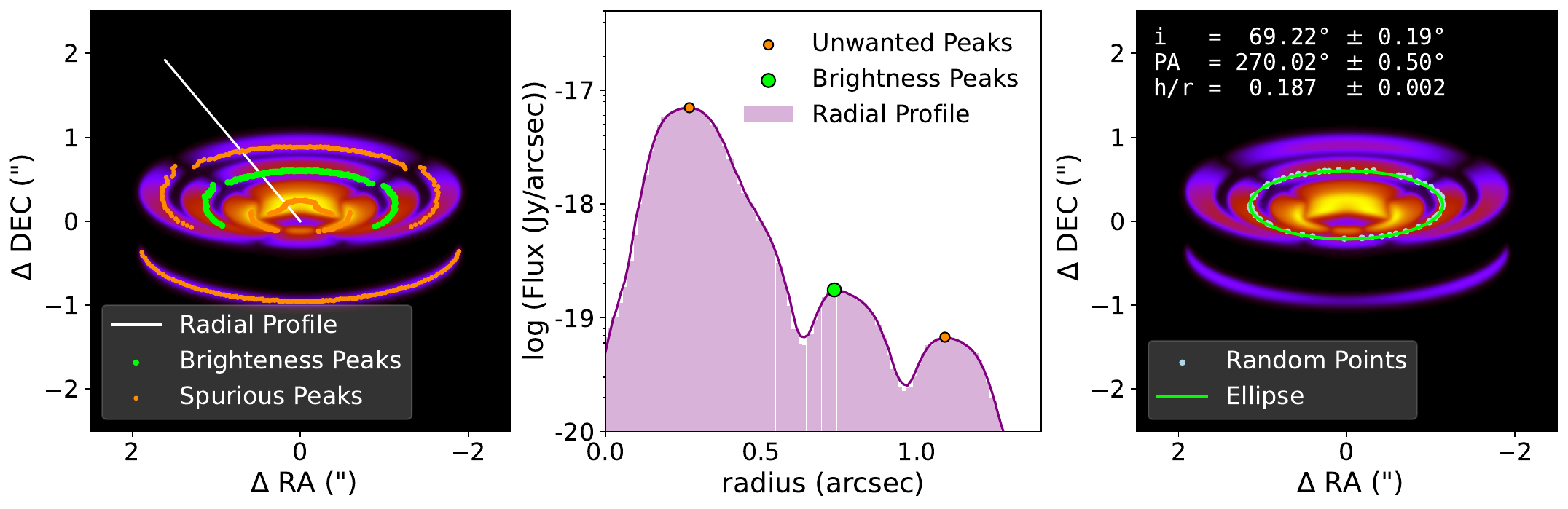}
    \caption{Illustration of the elliptical fit process on $Q_{phi}$ simulated disk image with multiple rings. Left: Disk image with a radial cut at one azimuthal angle in white. Points used for ellipse fitting are shown in green and unwanted deleted points in smaller orange dots. Middle: The radial profile extracted along the white line, where the purple curve corresponds to the smoothed profile. The maxima corresponding to the selected ring and the undesired maxima from other rings are shown in green and red, respectively. Right: Disk image with the fitted ellipse shown in white. A set of random points used to estimate uncertainties is plotted in blue.} \label{MaxPixFig}
\end{figure*}
{\tt{DRAGyS}} is designed to avoid the time-consuming step of model fitting to determine the disk structure necessary to extract the SPF. Instead, it estimates the disk geometry by assuming the disk is circular and by exploiting visible structures like rings that are used to delineate the position of the disk surface where scattering occurs. For geometrically thick disks, these elliptical rings are offset from the center, and the offset contains the information about the thickness and flaring of the disk surface \citep{lagage_anatomy_2006, avenhaus_disks_2018}.

{\tt{DRAGyS}} provides a comprehensive analysis of the disks, from their geometric parameter estimation to the SPF extraction. It follows three steps: i) Detection of the brightness peaks on the image, ii) Fitting by ellipses to estimate the disk geometry, and finally iii) Extraction of the SPF over the full disk or a restricted sector only, using the estimated geometric parameters.

We characterize the disk structure by the aspect ratio of the scattering surface, defined as the disk scattering surface height $h$ over the midplane radius $r$ from the star. Throughout our study, the power law exponent that governs the flaring of the disk scattering surface is $\chi$, and we refer to it as the scattering surface flaring exponent. We define the local height of the protoplanetary disk as:
\begin{equation} \label{eq_height}
\centering
    h(r) = \Href\left(\frac{r}{\rref}\right)^{\chi},
\end{equation}
where $h_\mathrm{ref}$ is the height of the disk at a distance $r_\mathrm{ref}$ from the star. The orientation of the disk is important to determine the scattering angles necessary to construct the SPF. So, {\tt{DRAGyS}} also fits for the inclination and the position angle of the disks. Together, these three parameters (respectively $h(r)$, $i$, $PA$) define the geometry of the disk surface (see Fig. \ref{DiskScheme}, right).

The method developed here is suitable for any disk observed in scattered light, in total and polarized intensity, at any wavelength. It is not limited by the geometric thickness of the disk, and can also be applied to optically thin disks. This approach requires the presence of well-defined concentric structures, such as rings. Most disks around T Tauri stars have a ring structure \citep{rapson_peering_2015, thalmann_resolving_2016, monnier_polarized_2017, bertrang_hd_2018, muro-arena_dust_2018, avenhaus_disks_2018, marel_protoplanetary_2019, benisty_optical_2022}. Incident starlight is attenuated by dust close to the central star. \citep{boer_multiple_2016, ginski_direct_2016}. Thus, we assume that maximum flux comes from the surface of the disk, illuminated by the star, and not at all or only partially affected by the inner rim of the ring. It is also necessary that instrumental effects (coronagraphic mask, imperfect PSF subtraction, low throughput, …) do not significantly alter the measured surface brightness.

\subsection{Step 1: identification of the ring structure}
The first step relies on image analysis and searches for maxima along radial profiles spaced by $1\degr$ in azimuth (see Fig. \ref{MaxPixFig}), enabling the location of the elliptical shape formed by the luminosity peaks within the ring from which the SPF is studied. Spatial and intensity filtering can be applied to optimize this detection, avoid confusion with other rings in the system, and eliminate spurious peaks due to weak local maxima in simulated images, or noise in observations (see section \ref{S4_RealData}). In addition, we perform a manual selection of peaks associated exclusively with the selected ring. Then, we down-select peaks that are spaced by a resolution element to ensure that the fitting is not biased by a cluster of correlated points, e.g., at the ellipse ansae. For this identification process and for SPF extraction, we do not use $r^2$-scaled image contrary to  \cite{stolker_scattered_2016, ginski_observed_2023} studies, as we do not have prior knowledge of the disk geometry. 

\subsection{Step 2: estimation of geometric parameters} \label{FittingPart}
The positions of the brightness peaks previously detected define an elliptical shape, whose parameters such as its center ($X_c, Y_c$), semi-minor ($m$) and semi-major ($M$) axis, and position angle ($PA$) can be estimated (see Fig. \ref{DiskScheme}, left). For this purpose, we use the least-square ellipse fitting method described in \cite{halir_numerically_1998}.

Then, the estimated ellipse parameters are transformed into disk geometrical parameters, taking into account the effects of projection onto the image plane (see Fig. \ref{MaxPixFig}). The inclination is determined by the ratio of the semi-minor and semi-major axis. The position angle $PA$ is measured in the usual way, from the horizontal left axis to the ellipse semi-minor axis corresponding to the disk near-side and counter-clockwise. Finally, the scattering surface height $h$, at a given midplane radius $r_\mathrm{mid} = M$, is defined as the distance $D$ between the estimated ellipse center and the central star, taking into account projection effects due to disk inclination\footnote{The distance $D$ is defined to take into account cases where the star and disk are not centered. To do this, we measure $D$ as the distance between the estimated semi-major ellipse and the central star.}. We define the aspect ratio of the scattering disk surface as $h/r = D/(r_\mathrm{mid}\sin{i})$.

Uncertainties associated with the disk geometric parameters are estimated by propagating the ellipse fitting error. Concretely, we compute the standard deviation, $\sigma$, of the relative distances between each maxima point and the fitted ellipse. From this standard deviation, we generate $N_p$ sets (typically $100$ sets) of points randomly distributed around the ellipse, following a normal distribution in radius using $r_\mathrm{mid}$ for the mean, and $\sigma$ for the standard deviation. The ellipse fitting process is applied to each of these sets to obtain a statistical distribution of geometric parameters. The uncertainty of each parameter is finally defined as the standard deviation of the estimated parameter values across the $N_P$ sets, providing a robust measure of the errors.

\subsection{Step 3: SPF extraction} \label{PhaseFunctionPart}
The SPF must be extracted from a restricted zone bordered by two ellipses located at the estimated disk surface. They are computed using estimated geometrical parameters, and follow the formulation from \cite{stolker_scattered_2016}. To accurately capture the flux from a ring, the extraction zone is limited by minimum and maximum midplane radii. This is necessary to exclude unwanted areas, such as gaps or other rings, which could alter the SPF, but also to focus the study on a well-defined part (radius) of the disk. This method therefore also allows to treat each rings separately, without assuming the same SPF throughout. This is particularly interesting for disks with multiple rings that could host different dust populations. It enables the analysis of the dust properties as a function of radius. In the same way, it enables to observe the radial dependence of light scattered on the surface of the disk by probing zones more or less distant from the center. Consequently, the extraction zone is chosen to be as wide as possible, within the ring observed. The influence of its position and width is discussed in Appendix \ref{APP:EZimpact}.

To construct the SPF, the flux is extracted for each pixel in this extraction zone, and the corresponding scattering angle $\Psi$ is computed as:
\begin{equation} \label{ScattAngle}
    \cos(\Psi) = \cos(\gamma) \cos(\phi)\sin(i) + \sin(\gamma)\cos(i),
\end{equation}
where $\gamma$ is the disk opening angle, which corresponds to the angle between the midplane and the disk scattering surface as seen from the star. 

To incorporate uncertainties from estimated geometric parameters, the SPFs are extracted over a fixed sampling in scattering angle, using $N_s$ different sets of parameters randomly taken at $1\sigma$ around their estimated value. This yields $N_s$ SPFs and their error given by the mean value and standard deviation of flux respectively in each bin. We average these SPFs to obtain the final curve, and the standard deviation of the mean gives the uncertainty linked to the geometric parameters. Lastly, we quadratically combine this uncertainty with the one from the flux extraction to obtain the total uncertainty of the extracted SPF.

\subsection{Features and requirements}
{\tt{DRAGyS}} is primarily designed to provide a complete analysis of ringed protoplanetary disks. It determines the disk geometry and extracts the SPF, covering all azimuths as well as each side separately to reveal eventual asymmetries. However, {\tt{DRAGyS}} can extract the SPF from any disks, including smooth ones, provided the disk surface geometry is known a priori (or modelled). it further includes several advanced options, relevant in specific cases. For example, it handles situations where the actual location of the star is offset from the center of the disk, and automatically corrects the extracted SPF from the limb brightening effect, whose details are discussed in section \ref{SPFBias}. The tool deals with both total and polarized intensity equally well. To simplify the rest of the paper, we will focus on polarized intensity results only.

\section{Benchmark tests with simulated data} \label{S3_SimulationPart}

\begin{table*}

\caption{Geometric fitting results from simulated disks.}
\centering
\begin{tabular}{cc|cccccc}
\toprule
\midrule
\multicolumn{2}{c|}{Inputted parameters} & \multicolumn{6}{c}{Estimated parameters}   \\ \midrule 
PA (deg)                    & i (deg)  & \multicolumn{3}{c|}{Position Angle (deg)}                                   & \multicolumn{3}{c}{Inclination (deg)}                  \\ 
                               &    &          R1       &          R2       & \multicolumn{1}{c|}{        R3       } &         R1       &         R2       &         R3       \\ \midrule              
                                     & 20 & 69.78 $\pm$ 3.31  & 70.88 $\pm$ 0.89  & \multicolumn{1}{c|}{70.48 $\pm$ 0.88}  & 18.92 $\pm$ 1.88 & 19.49 $\pm$ 0.33 & 20.14 $\pm$ 0.34 \\
              70               & 50 & 70.14 $\pm$ 0.82  & 69.71 $\pm$ 0.28  & \multicolumn{1}{c|}{69.89 $\pm$ 0.15}  & 49.09 $\pm$ 0.71 & 48.84 $\pm$ 0.15 & 48.77 $\pm$ 0.10 \\
                                     & 70 & 69.83 $\pm$ 1.72  & 69.65 $\pm$ 0.38  & \multicolumn{1}{c|}{69.86 $\pm$ 0.22}  & 68.05 $\pm$ 0.89 & 68.65 $\pm$ 0.13 & 68.82 $\pm$ 0.07 \\[1mm]
                                     & 20 & 201.51 $\pm$ 3.26  & 200.61 $\pm$ 0.93  & \multicolumn{1}{c|}{200.27 $\pm$ 0.86}  & 19.30 $\pm$ 1.44 & 18.84 $\pm$ 0.34 & 20.09 $\pm$ 0.33 \\
              200               & 50 & 199.28 $\pm$ 0.80  & 200.16 $\pm$ 0.27  & \multicolumn{1}{c|}{199.95 $\pm$ 0.19}  & 49.96 $\pm$ 0.64 & 48.93 $\pm$ 0.19 & 48.76 $\pm$ 0.12 \\
                                     & 70 & 199.43 $\pm$ 1.62  & 200.01 $\pm$ 0.62  & \multicolumn{1}{c|}{200.03 $\pm$ 0.19}  & 68.33 $\pm$ 0.95 & 68.81 $\pm$ 0.18 & 68.85 $\pm$ 0.04 \\[1mm]
                                     & 20 & 269.65 $\pm$ 2.17  & 269.99 $\pm$ 0.97  & \multicolumn{1}{c|}{269.83 $\pm$ 0.92}  & 19.34 $\pm$ 0.69 & 19.40 $\pm$ 0.32 & 20.13 $\pm$ 0.31 \\
              270               & 50 & 269.73 $\pm$ 0.70  & 270.12 $\pm$ 0.38  & \multicolumn{1}{c|}{269.84 $\pm$ 0.21}  & 49.12 $\pm$ 0.80 & 49.76 $\pm$ 0.35 & 49.01 $\pm$ 0.13 \\
                                     & 70 & 269.45 $\pm$ 1.95  & 269.81 $\pm$ 0.32  & \multicolumn{1}{c|}{270.03 $\pm$ 0.30}  & 68.29 $\pm$ 1.37 & 68.89 $\pm$ 0.11 & 69.23 $\pm$ 0.09 \\[1mm]
                                     & 20 & 299.25 $\pm$ 3.30  & 299.44 $\pm$ 0.81  & \multicolumn{1}{c|}{300.12 $\pm$ 0.76}  & 19.09 $\pm$ 1.30 & 19.16 $\pm$ 0.33 & 20.01 $\pm$ 0.33 \\
              300               & 50 & 299.81 $\pm$ 1.29  & 299.97 $\pm$ 0.21  & \multicolumn{1}{c|}{299.70 $\pm$ 0.16}  & 49.12 $\pm$ 0.75 & 48.97 $\pm$ 0.14 & 48.84 $\pm$ 0.11 \\
                                     & 70 & 299.52 $\pm$ 2.75  & 300.08 $\pm$ 0.33  & \multicolumn{1}{c|}{299.94 $\pm$ 0.25}  & 69.13 $\pm$ 1.00 & 68.93 $\pm$ 0.09 & 69.03 $\pm$ 0.06 \\
                                     \hline
\label{FittingTableMCFOST}
\end{tabular}\tablefoot{Comparison between the input inclination $i$ and position angle $PA$ in the {\tt{MCFOST}} models and those estimated by our disk geometry fitting process for the 3 rings R1, R2, and R3.}
\end{table*}

\begin{figure}
\centering
\includegraphics[width=\hsize]{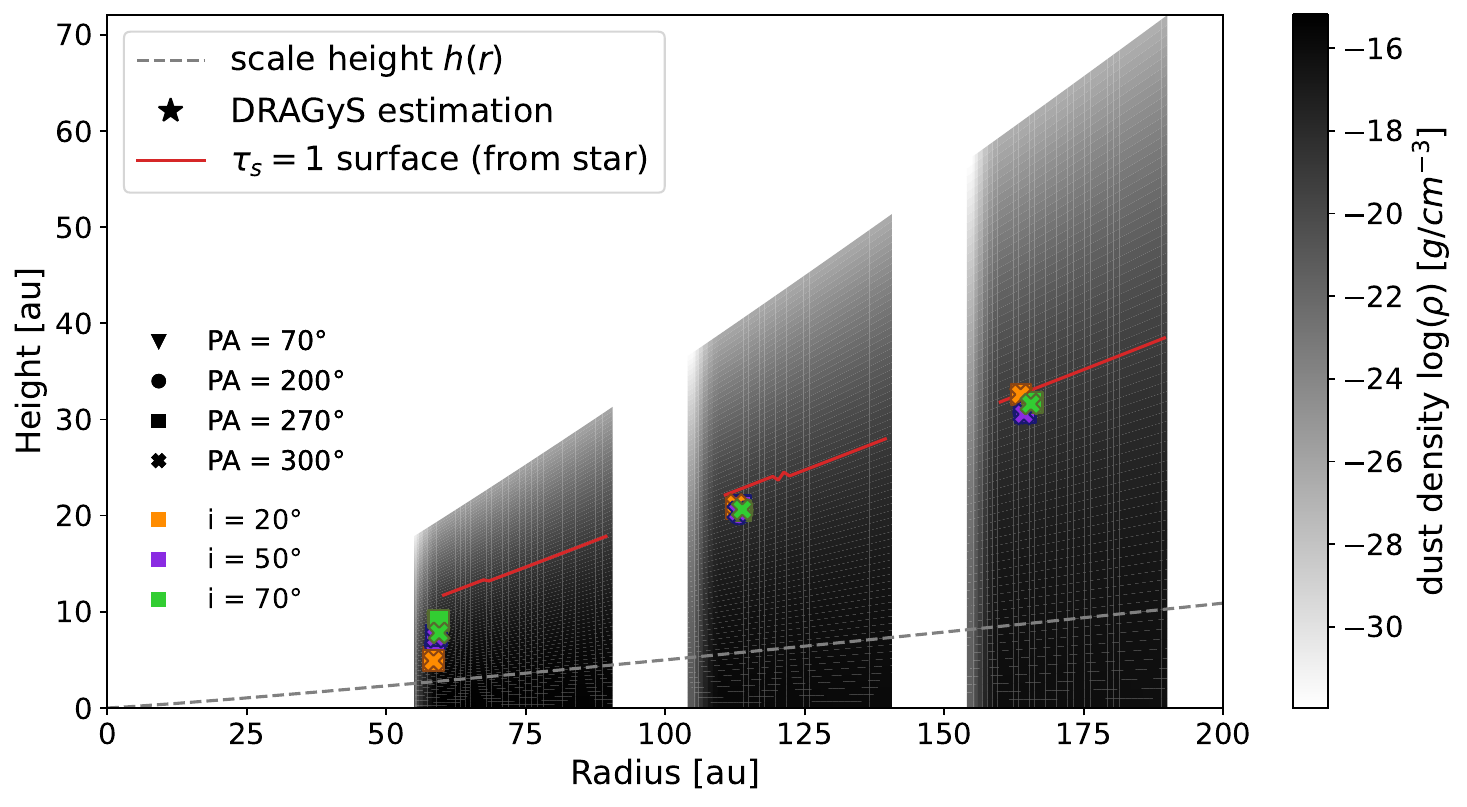}
  \caption{Comparison of estimated scattering height estimation with the $\tau_s=1$ surface integrated from the star (red) for each ring (respectively between $60$-$90$ au, $110$-$140$ au, and $160$-$190$ au). We overplot the scattering height surface estimated using {\tt{DRAGyS}}, for each inclination and position angle, in different colors and markers. We add the dust density map in the background and, we also plot the gas pressure scale height in dashed gray curve, using the input {\tt{MCFOST}} parameters.}
     \label{Fig:Tau_Surface}
\end{figure}
\begin{figure*}
    \centering
    \includegraphics[width=\hsize]{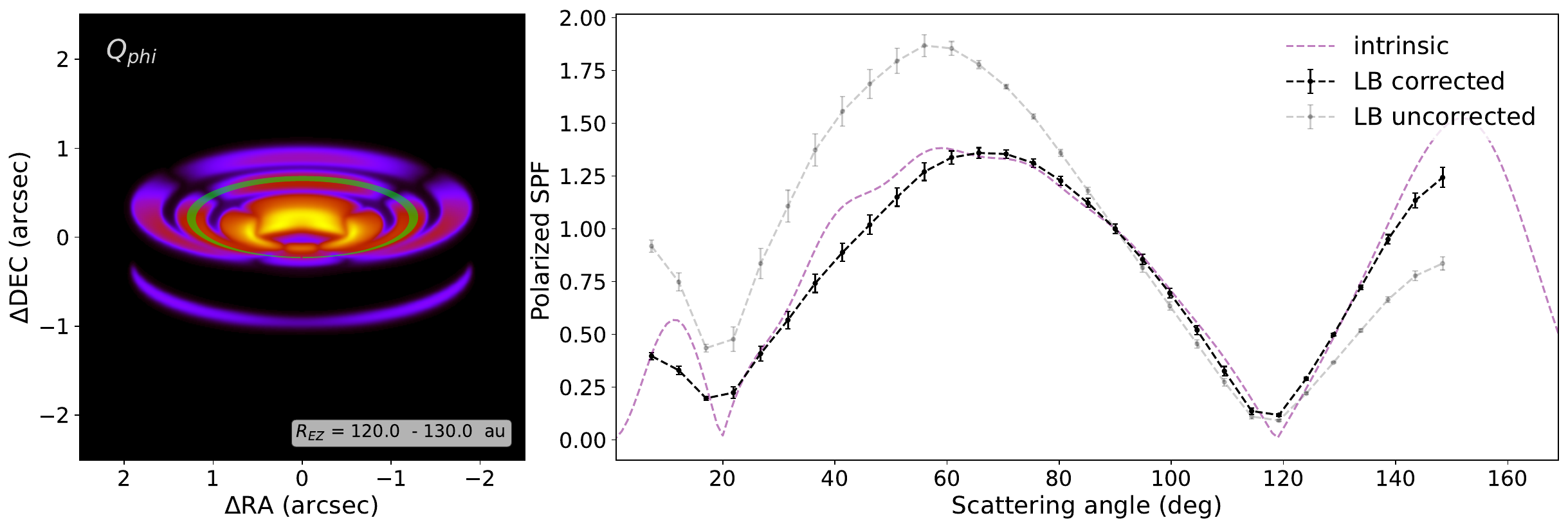}
    \caption{Typical case of SPF extraction for simulated data. Using a disk with $PA = 270$\degr~and $i = 70$\degr. 
    Left: Disk image in polarized intensity, with an extraction zone defined by the green-shaded annulus. Right: Polarized SPF normalized at $90$\degr~ directly extracted from the disk surface (gray) and corrected from the limb brightening effect (black). The theoretical intrinsic SPF is plotted in dashed purple line.}
    \label{SPFSimul}
\end{figure*}
\subsection{Making a mock observational image}
As a first test, we apply {\tt{DRAGyS}} to simulated images of protoplanetary disks. These simulations are performed using the 3D radiative transfer code {\tt{MCFOST}} \citep{pinte_monte_2006, pinte_benchmark_2009}. This software follows the Stokes formalism \citep{bohren_absorption_1983} to generate $I$, $Q$, and $U$ images of the disk. The Stokes formalism allows to follow the evolution of a light beam through successive scattering events. On top of the disk images, {\tt{MCFOST}} also outputs the theoretical SPF expected from the dust distribution used in the model. This allows a direct comparison between the SPF extracted from a simulated disk image and the theoretical SPF computed by {\tt{MCFOST}} (see Fig.\ref{SPFSimul}, right). 

Synthetic disks are located at a distance of $100$~pc and composed of 3 rings at $60$-$90$ au, $110$-$140$ au, and $160$-$190$ au. They are composed of silicate grains \citep{draine_optical_1984} with a power-law size distribution $n(a)da = a^{-p}da$ ranging from 0.005 to 3 $\mu m$ with a power-law index $p = 3.5$ \citep{dohnanyi_collisional_1969}. Although these grains may not be representative of those expected in protoplanetary disks, they produce a complex SPF with multiple minima and maxima, placing maximum stress on the method. This is well suited to test {\tt{DRAGyS}} and evaluate its strengths, biases, and weaknesses. We vary parameters such as position angle ($PA=70$\degr, $200$\degr, $270$\degr, $300$\degr) and inclination ($i=20$\degr, $50$\degr, $70$\degr) to test the robustness of this method. The gas pressure height is defined as $h_\mathrm{gas}=5~\mathrm{au}(r/100~\mathrm{au})^{1.125}$. The density profile follows a power-law, $\Sigma(r) \propto r^{-p}$, with surface density exponent $p = 0.5$. The dust mass is fixed to $M_d = 10^{-4} M_\odot$ in the fiducial model. All computed $Q_{phi}$ images \citep{schmid_limb_2006} are obtained at $1.6~\mu$m, and convolved using VLT/SPHERE instrument diameter We display all simulated $Q_{phi}$ images on appendix \ref{APP:Models}.

\subsection{Geometric parameters fitting} \label{Sect_FitSimu}
We directly compare the input model values with the estimated inclination, position angle and scattering height values extracted with {\tt{DRAGyS}}. Models are composed of 3 rings, which also enable an analysis of potential dependence on the position of the ellipse used for estimation. The estimated parameters for each ring are listed in Table \ref{FittingTableMCFOST}. 

An overall comparison shows that parameter estimation tends to improve with increasing radius. Analysis of the inner ring shows larger discrepancies with reference values, both for position angle and inclination, mainly due to the smaller number of pixels available for fitting when looking at the inner ring. For the same reason, measurement errors decrease with increasing radius. Specifically, our method recovers the inclination within $\sim2$\degr~of the theoretical value for all cases. The worst-case error is for the inner ring estimation of the $70$\degr~inclined-disk. We also observe an underestimation of inclination of slightly less than $1$\degr~on average. In terms of position angle, the estimated value nearly always includes the theoretical value within its confidence range, with a maximum offset of $\sim1.5$\degr, when analyzing a weakly inclined inner ring. Regarding the measured scattering height, we compare it with the $\tau_s=1$ surface as computed from the star. This surface corresponds to the last scattering event that emerges from the disk (see fig \ref{Fig:Tau_Surface}). For very optically thick and flared disks, as is usually the case for protoplanetary disk in the optical and near-infrared, the photons reaching the observers come mainly from the disk surface and this surface is well estimated by this $\tau_s=1$ surface. See Appendix \ref{APP:DustMass_height} for a comparison of the measured surface versus the $\tau_s=1$ surface for various disk masses. For the inner ring, the estimations are all lower in the disk than the $\tau_s=1$ surface, and more spread out. This is due to the direct illumination of the inner rim, combined with the decreasing number of points available for ellipse fitting when approaching the central star. Our method reaches its limits as the distance between the projected stellar position and the ellipse center becomes smaller and thus harder to measure precisely. Comparing the estimated scattering heights with the pressure scale height used by {\tt{MCFOST}} gives a mean ratio of $\sim 3.7$, within the same range as found in previous studies on different targets \citep{chiang_spectral_2001, avenhaus_disks_2018, ginski_direct_2016}. We discuss this comparison for several disk dust masses in appendix \ref{APP:DustMass_height}.

For these tests, uncertainties are based on noiseless simulated data. They represent a minimum uncertainty intrinsic to the parameter estimation method. In other words, the estimated geometric parameter uncertainties are optimistic relative to the case of a real dataset.

\subsection{Biases in extracted SPF} \label{SPFBias}
We compare the "true" (intrinsic) SPF from the model, with the SPF directly extracted from the disk surface (see Fig. \ref{SPFSimul}). The extracted SPF has the correct qualitative behavior, with troughs at the same scattering angles and peaks roughly at the correct position. The shift in the maximum of the extracted SPF is due to two major effects that modify the shape of the SPF. 

The first one is multiple scattering, which is characterized by photons that undergo several scattering events on the dust grains inside the protoplanetary disk before reaching the surface and leaving the disk. \citet{tazaki_fractal_2023} showed that multiple scattering affects the polarized SPF of protoplanetary disks, particularly when the single scattering albedo of grains is large. There is no simple prescription to correct for this effect, since it depends on the disk structure and optical depth.

The second one is the limb brightening effect, which results in the near side of the disk appearing brighter than the far side. In the context of optically thick disk observations, \cite{tazaki_fractal_2023} showed that this purely geometric effect is caused by the optical depth, which varies according to the location on the inclined disk. In other words, the optical depth measured from the surface to the observer does not reach unity at the same height in the disk over all azimuths, which significantly influences the intensity received from different surface parts, and then, on the measurement of the SPF. The method developed here estimates, and can correct SPF for, the limb brightening effect, following the formulation from \cite{tazaki_fractal_2023}:
\begin{equation} \label{eq_LB}
    \mathrm{LB} = \frac{\cos(\gamma) \sin(\gamma' - \gamma)}{\cos(\gamma)\sin(\gamma' - \gamma) + \cos(i) \cos(\gamma' - \gamma) - \sin(\gamma') \cos(\Psi)}, 
\end{equation}
where $\gamma = h/r$, and $\gamma' = dh/dr = \chi h/r$.

\begin{figure}
\centering
\includegraphics[width=\hsize]{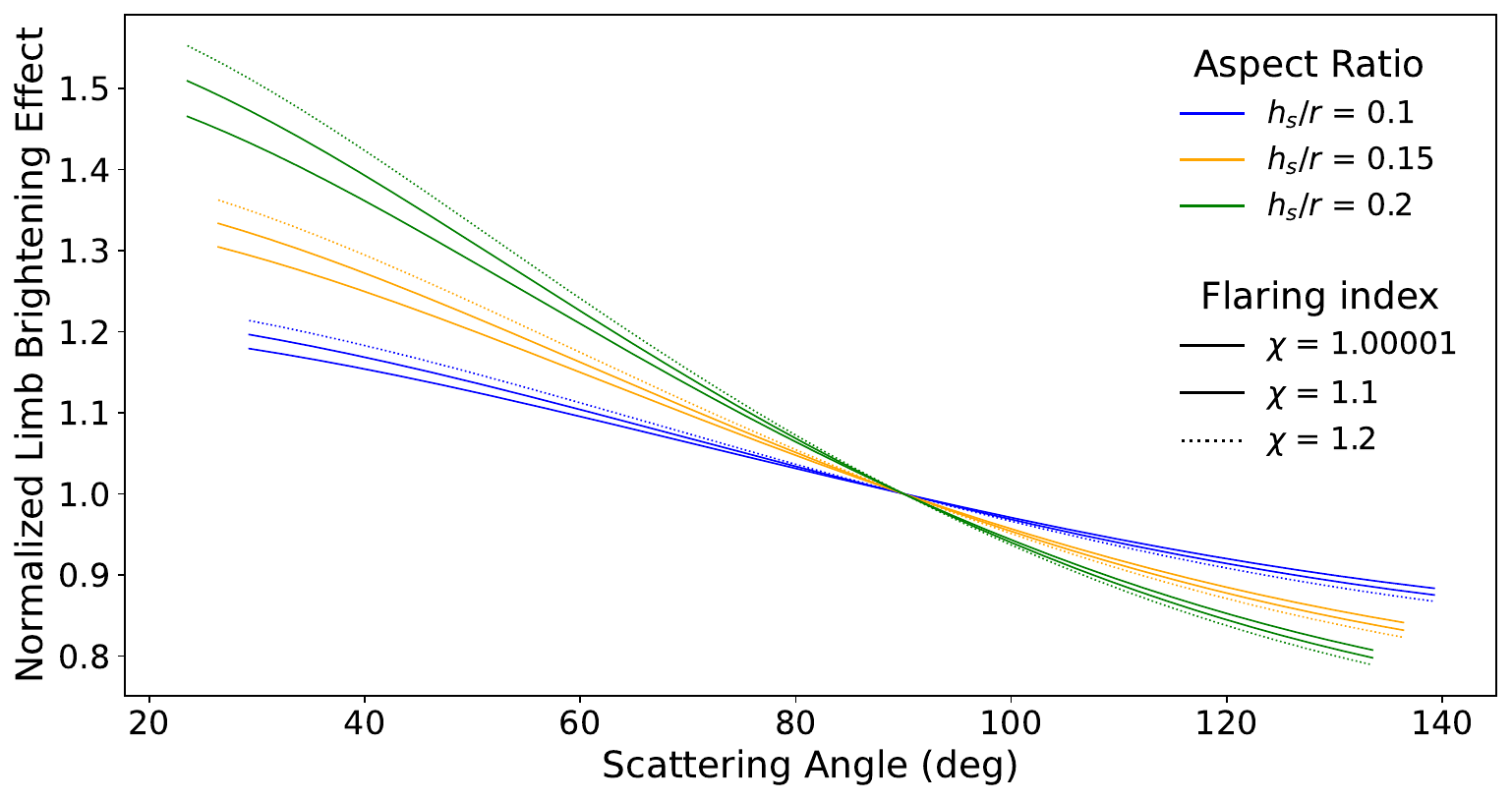}
  \caption{Limb brightening effect for different disk geometry. The forward scattering part is more affected than the backward scattering.}
     \label{LimbBrightening}
\end{figure}

The limb brightening effect plotted in figure \ref{LimbBrightening} shows that there is only a slight dependence on the disk flaring exponent. On the contrary, limb brightening is dominated by the aspect ratio. We therefore assume a fixed flaring exponent value close to unity for the {\tt{DRAGyS}} SPF extraction process and for the limb brightening effect estimation. Uncertainties of geometrical parameters are propagated through the limb brightening correction process. Figure \ref{SPFSimul} shows that the limb brightening-corrected SPF is a much better match to the expected SPF from the model.

\section{Application to observational data} \label{S4_RealData}
\begin{figure}
    \centering
    \includegraphics[width=\hsize]{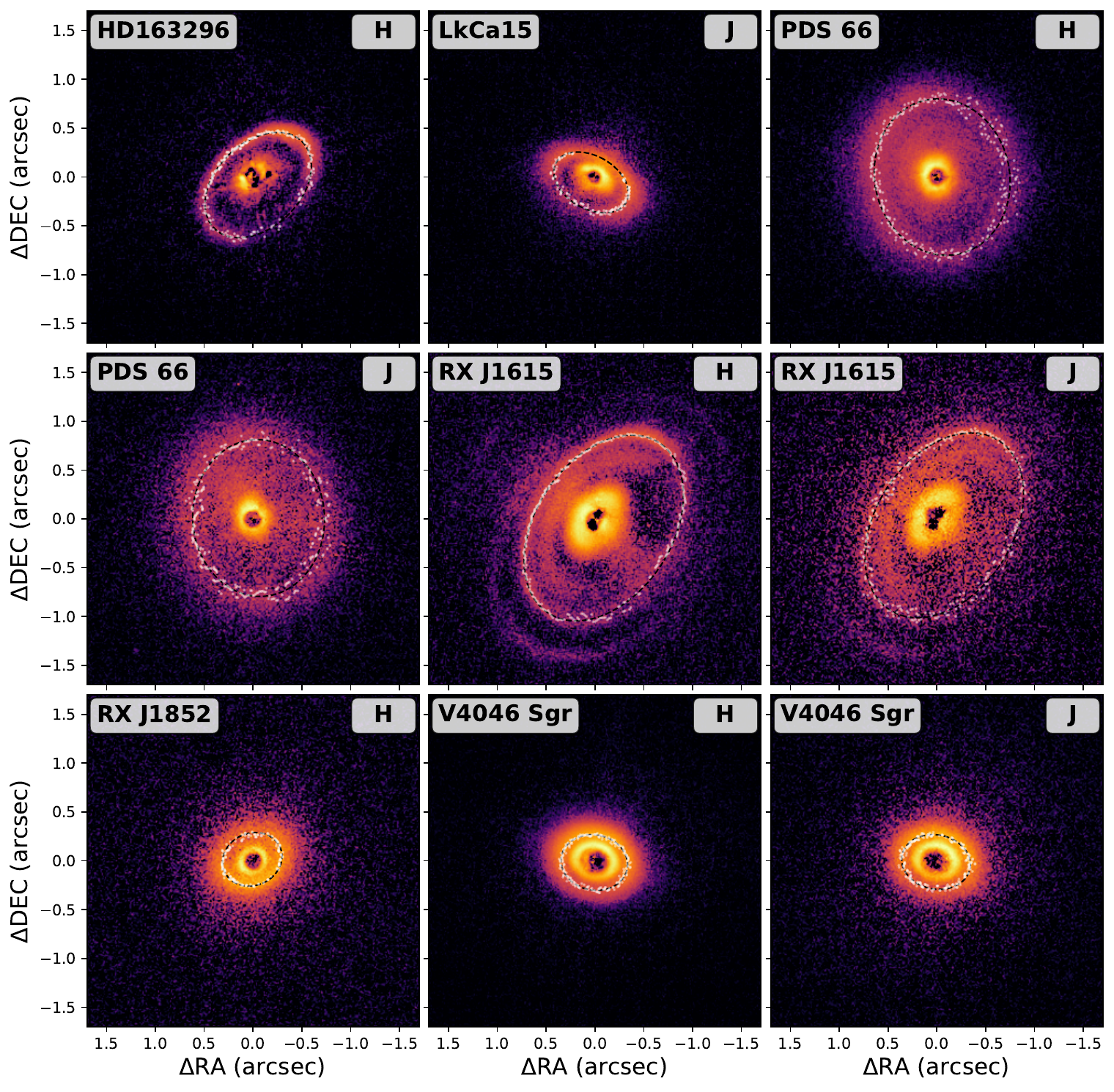}
    \caption{$Q_\phi$ images of 6 protoplanetary disks observed in J or H bands using DPI with the VLT/SPHERE IRDIS instrument. The maximum pixel positions for parameter estimation are shown as white dots. and fitted ellipse in black dashed line}
    \label{RealObservation}
\end{figure}
\begin{table*}
\caption{Geometric fitting values for VLT/SPHERE observations using {\tt{DRAGyS}}.}
\label{ObservationTable}
\parbox{\textwidth}{
\centerline{\begin{tabular}{@{\extracolsep{4.0mm}} l c c c c c c c}
\toprule\midrule
 Object    & Band & distance ($pc$) &         $PA$ (°)       &       $i$ (°)     &     $h/r$         &  $R_{in}$ (au) & $R_{out}$ (au) \\ \midrule
 HD 163296 &  H   &      101.1      &   134.22 $\pm$ 1.19    &  46.77 $\pm$ 0.87 & 0.210 $\pm$ 0.009 &       55      &        75       \\
           &      &                 &        (134.8)         &        (46.0)     &        (0.214)    &                                 \\[1mm]   
 LkCa 15   &  J   &      157.2      &   60.18  $\pm$ 1.84    &  48.93 $\pm$ 1.08 & 0.220 $\pm$ 0.011 &       58      &        94       \\
           &      &                 &          (60)          &         (50)      &        (0.181)    &                                 \\[1mm]   
 PDS 66    &  H   &      97.9       &   191.98 $\pm$ 4.76    &  30.93 $\pm$ 2.48 & 0.128 $\pm$ 0.025 &       64      &       104       \\
           &  J   &      97.9       &   183.28 $\pm$ 4.24    &  32.84 $\pm$ 2.27 & 0.133 $\pm$ 0.023 &       64      &       104       \\
           &      &                 &        (189.2)         &       (30.3)      &       (0.138)     &                                 \\[1mm]  
 RX J1615  &  H   &      155.6      &   147.87 $\pm$ 0.96    &  47.47 $\pm$ 0.65 & 0.178 $\pm$ 0.007 &      146      &       181       \\
           &  J   &      155.6      &   147.18 $\pm$ 1.18    &  47.19 $\pm$ 0.81 & 0.132 $\pm$ 0.008 &      146      &       181       \\
           &      &                 &        (145.0)         &        (47.2)     &       (0.168)     &                                 \\[1mm]   
 RX J1852  &  H   &      147.1      &   293.10 $\pm$ 4.38    &  30.87 $\pm$ 2.50 & 0.106 $\pm$ 0.024 &       36      &        68       \\
           &      &                 & (304\tablefootmark{a}) &      (30)         &       (0.067)     &                                 \\[1mm]   
 V4046 Sgr &  H   &      71.5       &   79.01  $\pm$ 3.99    &  34.43 $\pm$ 2.09 & 0.096 $\pm$ 0.020 &       23      &        34       \\
           &  J   &      71.5       &   79.04  $\pm$ 3.14    &  38.40 $\pm$ 1.83 & 0.062 $\pm$ 0.017 &       23      &        34       \\
           &      &                 &        (74.7)          &        (32.2)     &       (0.126)     &                                 \\ \midrule
\end{tabular}}}
\tablefoot{We display the parameters used to define our extraction zone to extract the SPF from the surface of protoplanetary disks using {\tt{DRAGyS}}. To ensure the most accurate comparison, the target distances and the radii for the extraction zone $R_{in}$ and $R_{max}$ are taken from \cite{ginski_observed_2023}, as well as the values of the geometric parameters in parentheses.\\
\tablefoottext{a}{The PA of RX J1852 is set to $124$° on \cite{ginski_observed_2023} paper, but we adopt the value of $PA+180$\degr~as required to match the direction of forward scattering.}}
\end{table*}
Several disks have clearly visible rings for which the SPF has already been estimated in the past \citep{ginski_observed_2023}, which is useful to test the {\tt{DRAGyS}} extraction process. Here we focus on 6 protoplanetary disks observed in scattered light with the Spectro-Polarimetric High-contract Exoplanet REsearch facility (SPHERE \cite{beuzit_sphere_2019}) at VLT, using the IRDIS near-infrared camera \citep{dohlen_notitle_2008}. To obtain a polarized intensity image of the protoplanetary disk, the instrument is operated in dual-beam polarization (DPI) mode and observations are made in the J and/or H bands (see Fig. \ref{RealObservation}). It should be noted that although the data are obtained in the classic Stokes Q and U formalism, we use the azimuthal Stokes Q images, $Q_{\phi}$, to carry the SPF extraction from each observation \citep{schmid_limb_2006}.

Geometric fitting of protoplanetary disks from real observations presents a few challenges. The presence of noise complicates the identification of true intensity maxima, so additional steps must be applied to the fitting process compared with the method presented in Section \ref{S2_GeoFit}. These consist essentially in a spatial averaging of the image and a more aggressive filtering on the intensity maxima detection. 

We list the parameters estimated with {\tt{DRAGyS}} in Table \ref{ObservationTable} and maps of points detected for ellipse fitting are shown in Figure \ref{RealObservation}. The comparison between the reference geometrical parameters and those estimated with {\tt{DRAGyS}} yields consistent results in most cases. Larger deviations in PA are sometimes found, especially when the disk inclination is low, for example in PDS 66 (J-band), V4046 Sgr (H-band), and RXJ 1852 at worst. For the aspect ratio, some estimations are consistent between our method and the reference data, but there can be large discrepancies, as in the case of RX J1852 or LkCa~15. Aspect ratio estimations also vary from band to band. Therefore, the best way to compare our estimation is to observe the extracted SPF.

\begin{figure*}
    \centering
    \includegraphics[width=.9\textwidth]{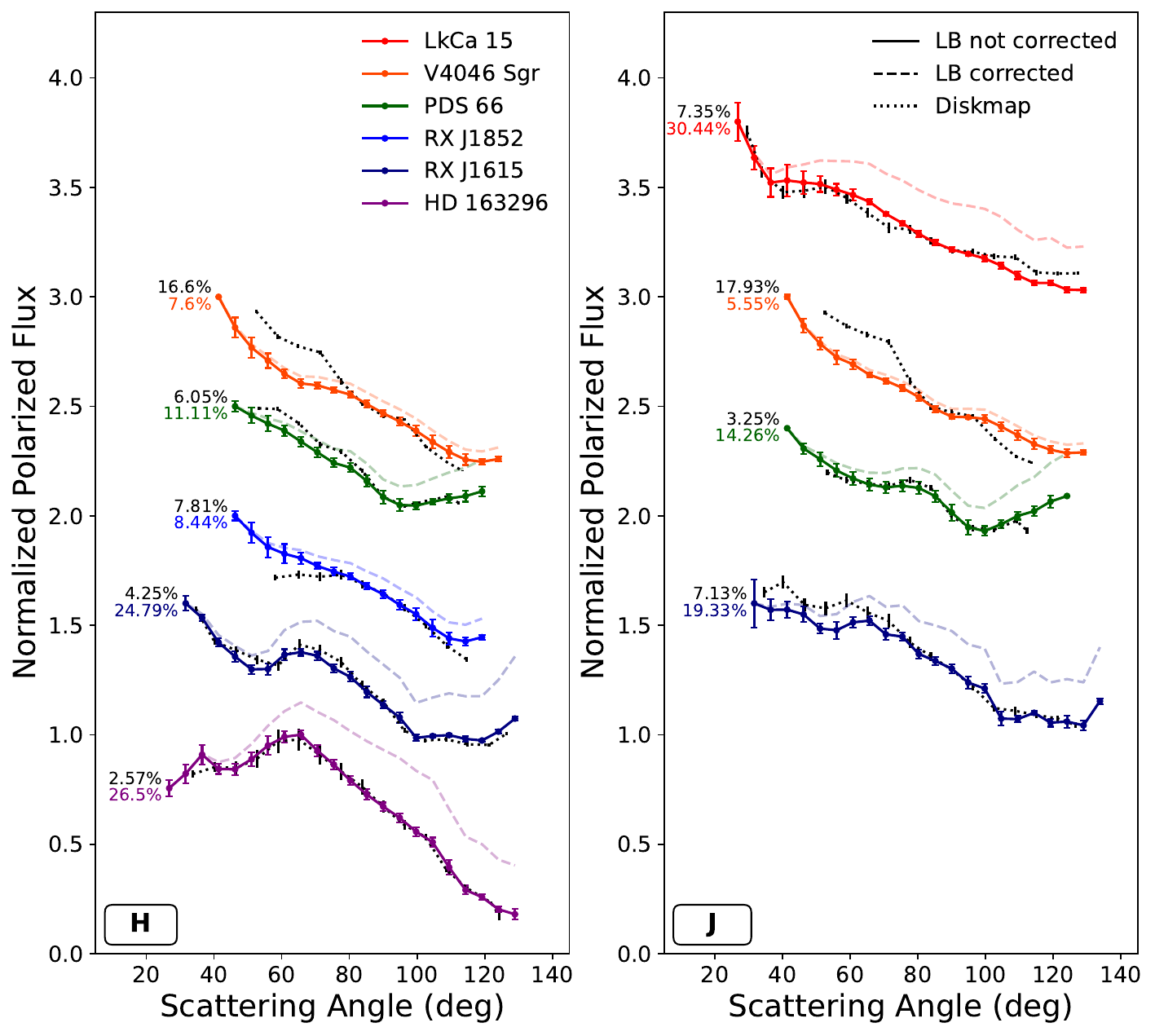}
    \caption{Polarized SPFs extracted from SPHERE $Q_{phi}$ observations \citep{ginski_observed_2023} of protoplanetary disks. The left and right panels are for $H$ and $J$ band observations, respectively. Colored solid and dashed lines are respectively SPFs uncorrected and corrected from the limb brightening effect, extracted using {\tt{DRAGyS}} with fitted geometric parameters. Black dotted lines show the reference SPFs from {\tt{Diskmap}} \citep{stolker_scattered_2016} following the geometric parameters from \cite{ginski_observed_2023}. Limb brightening uncorrected and {\tt{Diskmap}} SPFs are normalized to $90$\degr. We normalize the corrected SPF to the minimum scattering angle in common with the uncorrected one, emphasizing the slope change induced by the limb brightening correction. 
    An offset is added between each curve for better rendering. The percentage of mean offset between SPF before and after removing the limb brightening are shown in color, and the ones between {\tt{DRAGyS}} (without LB correction) and {\tt{Diskmap}} are in black.}
    \label{DiskmapVSDRAGyS}
\end{figure*}

The geometric parameters estimated in Table \ref{ObservationTable} are applied to extract the SPF from an extraction zone on the disk surface using {\tt{DRAGyS}} (See Fig. \ref{DiskmapVSDRAGyS}). We compare our results with reference SPFs using {\tt{Diskmap}} and reference geometries from \cite{ginski_observed_2023}, keeping the same width and position of the extraction zone for both processes. The difference between the two SPFs is quantified using the relative error over the common interval in the scattering angle for both SPFs.

Both methods cover very similar scattering angle ranges. As a result, we typically obtain good agreement in terms of SPF, with differences in the range $2\%$ to $18\%$. As with the geometrical parameter estimation, the most problematic case remains V4046 Sgr, where scattering angle ranges and shapes are not consistent with {\tt{Diskmap}}. This system presents the largest difference ($\sim17\%$ in both band), in line with the fact that its geometric estimation was the worst in the sample. If the defined geometry is different, SPFs will usually be different. In contrast, a much better match is observed in the case where the geometries are consistent for both methods, such as for RX~J1615,  HD~163296 (H-band) and PDS~66 (J-band), with a percentage difference of $2.57\%$, $4.25\%$ and $3.25\%$ respectively. These two extremes demonstrate the significant influence of the estimated geometry on the SPF, but also the consistent SPF produced by the two methods for similar geometries.

Now, consider the SPF corrected for limb brightening, which is estimated with {\tt{DRAGyS}} (see Fig \ref{DiskmapVSDRAGyS}). The general shape of the curves does not significantly change, with troughs and peaks still present at the same scattering angles, but the slope becomes shallower in all the cases considered here. This deviation may vary according to the disk geometry: it can be negligible in some cases, such as for V4046 Sgr and RX~J1852 in the J and H bands, and strong enough to significantly modify the SPF in other cases, as occurs for LkCa~15, RX~J1615, and HD163296. Moreover, we also compute the percentage difference between SPF with and without correction. Limb brightening typically impacts SPF by about $10\%$ on average, and can even reach $\sim 30\%$ in some cases. It is therefore important to correct for it before interpreting SPFs.

In summary, the method described here not only provides a relatively accurate estimation of SPF, but also corrects an effect that can have a significant amplitude depending on the observations, which was not corrected in previous results. We discuss this point in section \ref{S5_Discussion}.

\section{Discussion} \label{S5_Discussion}
\subsection{Limiting factors}
The key factor for SPF extraction is the positioning of the extraction zone. A correct positioning ensures accurate coverage of the disk scattering surface (here, the ring). Incorrectly estimated geometric parameters lead to a mismatch between the extraction zone and the ring. In the worst case, this can lead to the inclusion of zones outside the disk surface, or coming from another ring, leading to a wrong SPF.

For simulated disk images, with known inclinations and PA, {\tt{DRAGyS}} finds a systematic underestimation of $\sim 1$\degr~for the inclination, but the correct PA. The exact value depends on the disk model, but also on the manual filtering of the intensity maxima (see section \ref{Sect_FitSimu}). Obviously, it's harder to estimate the position angle for a weakly inclined disk, whereas a highly inclined disk makes inclination estimation more difficult, and even more if combined with a large flaring. For VLT/SPHERE images, the true inclination values are unknown. {\tt{DRAGyS}} recovers parameters similar to those found by \citet{ginski_direct_2016}.

The extracted SPF depends on the intrinsic SPF, limb brightening, and multiple scattering effects. The limb brightening effect can amplify the flux received at low scattering angles and must be corrected for. However, multiple scattering remains an uncalibrated factor. It can introduce a significant bias into the polarized SPF, especially since multiple scattering leads to a reduction in net polarization fraction due to interactions between light and dust grains, especially at small scattering angles \citep[see Fig. 9 in][]{tazaki_fractal_2023}. This effect can be evaluated a posteriori using radiative transfer models. While desirable, the development of a method to quantify this effect and obtain a closer-to-reality SPF is beyond the scope of this paper. For the simulation described here (Fig. \ref{SPFSimul}), the extracted SPF is almost fully consistent with the intrinsic SPF of dust grains. However, there remains the multiple scattering effect, which depends on dust properties that have only been partially explored in our simulations. To further assess the impact of this effect, we extract SPFs from disks with different dust compositions and grain size distributions in Appendix \ref{APP:SPF_dust_prop}. Overall, we find that multiple scattering only has a modest influence on SPFs across the tested conditions.

As these two effects of limb brightening and multiple scattering have opposite influences on polarized SPF, they may counterbalance each other. These findings underscore the importance of considering both geometric and physical factors when interpreting SPFs.

\subsection{Achievements and highlights}
Despite these limitations, {\tt{DRAGyS}} accurately retrieves the disk geometric parameters on both simulated and telescopic data. It allows a fast and reliable extraction of SPFs without prior knowledge of disk geometry.

The accuracy of SPF recovery depends on the correct positioning of the extraction zone over the disk surface. It can be affected if the zone is not well matched to the ring position and width (see Appendix \ref{APP:EZimpact}).

High-inclination disks are particularly sensitive to errors in the extraction zone position. However, a modest error on inclination ($\pm 2$\degr), or on the aspect ratio (typically $0.025$), does not significantly alter the extracted SPF on this type of disk (see Appendix \ref{APP:ParamsImpactSPF}). The simulations performed here are focused on disks with narrow, closely-spaced rings, an extreme scenario that demonstrates the robustness of the tool on complex structures.

Concerning the advantages of the method, {\tt{DRAGyS}} allows extracting the SPF in total and polarized intensity from the protoplanetary disk surface, avoiding long model-fitting computations per disk, thanks to a geometric fitting approach directly on the disk images. In addition, the method presented here enables the complete study of a wide range of disks from geometric fitting on disk image to SPF extraction and limb brightening correction. The only constraint is that a structure like a ring is required for the geometrical fitting process only. But it is also possible to manually define geometrical parameters, by-pass that part of the fitting process, and apply the SPF extraction directly for disks with already known geometry.

A capability of {\tt{DRAGyS}} that we have not used here, but that will prove powerful in the future is the possibility to separate the extracted SPF into two zones, typically on either side of the minor axis of the disks. Indeed, observations often reveal disks with brightness asymmetries on their surface. These could be due to an asymmetric dust density distribution around the disks, but also because of shadows cast by a misaligned inner disk, for example. These asymmetries can therefore bias the inferred dust properties and {\tt{DRAGyS}} provides an automated process to identify systems with significant asymmetry, and allows the user to extract the SPF from a subset of the data that is deemed uncontaminated by intrinsic asymmetries.

We explored the SPFs extracted for all disks considering each half separately (see Appendix \ref{Apx4_allSPF}). This is particularly interesting for sources HD163296 and RX J1615, where the brightness ratio between the disk ansae is around 2.75 and 1.75 respectively. This ratio is close to unity for the other sources, which does not have a noticeable impact on the SPFs.

\section{Conclusion} \label{S6_Conclusion}
Here, we present {\tt{DRAGyS}}, a new method to extract the SPF by analyzing a disk from geometric parameters estimation to SPF extraction. {\tt{DRAGyS}} represents a major advance in the study of the SPF extracted from ring-shaped protoplanetary disks. Unlike previous methods requiring prior modeling of the disk, this tool directly uses images to estimate the geometry and extract the SPF, while correcting for geometric effects such as limb brightening, which are often neglected but can significantly influence the shape of the SPF.

Tests on synthetic images have shown that the geometric parameters of the disk, such as inclination and position angle, can be estimated within $1$\degr~accuracy, although minor biases may occur, particularly in the case of disks with high inclination or more complex structures. These estimations are made by detecting an elliptical shape created by brightness peaks on the disk image. Application to real data from polarized near-infrared observations demonstrates the robustness of the tool, which provides consistent results with existing techniques such as {\tt{Diskmap}} while offering advantages in terms of speed and execution simplicity. Therefore, {\tt{DRAGyS}} is particularly suitable for studies of large disk samples.

Finally, this tool also has the potential to improve the characterization of asymmetric or complex disks, by enabling the separate extraction of SPF for different regions of the disk. The correction of limb brightening, combined with a detailed analysis of surface asymmetries, could open up new perspectives in the study of dust grain properties and planet formation in protoplanetary disks.

\section*{Data availability}
The {\tt{DRAGyS}} tool is available in GitHub repository: \url{https://github.com/mroumesy/DRAGyS}.

\begin{acknowledgements}
We would like to thank the anonymous referee whose intuition led to a better paper. This project has received funding from the European Research Council (ERC) under the European Union’s Horizon Europe research and innovation program (grant agreement No. 101053020, project Dust2Planets, PI: F. M\'enard).
\end{acknowledgements}

\bibliography{Biblio_V2}

\onecolumn
\begin{appendix}
\section{MCFOST disk models for testing DRAGyS} \label{APP:Models}
\begin{figure}[htbp]
    \centering
    \includegraphics[width=0.9\textwidth]{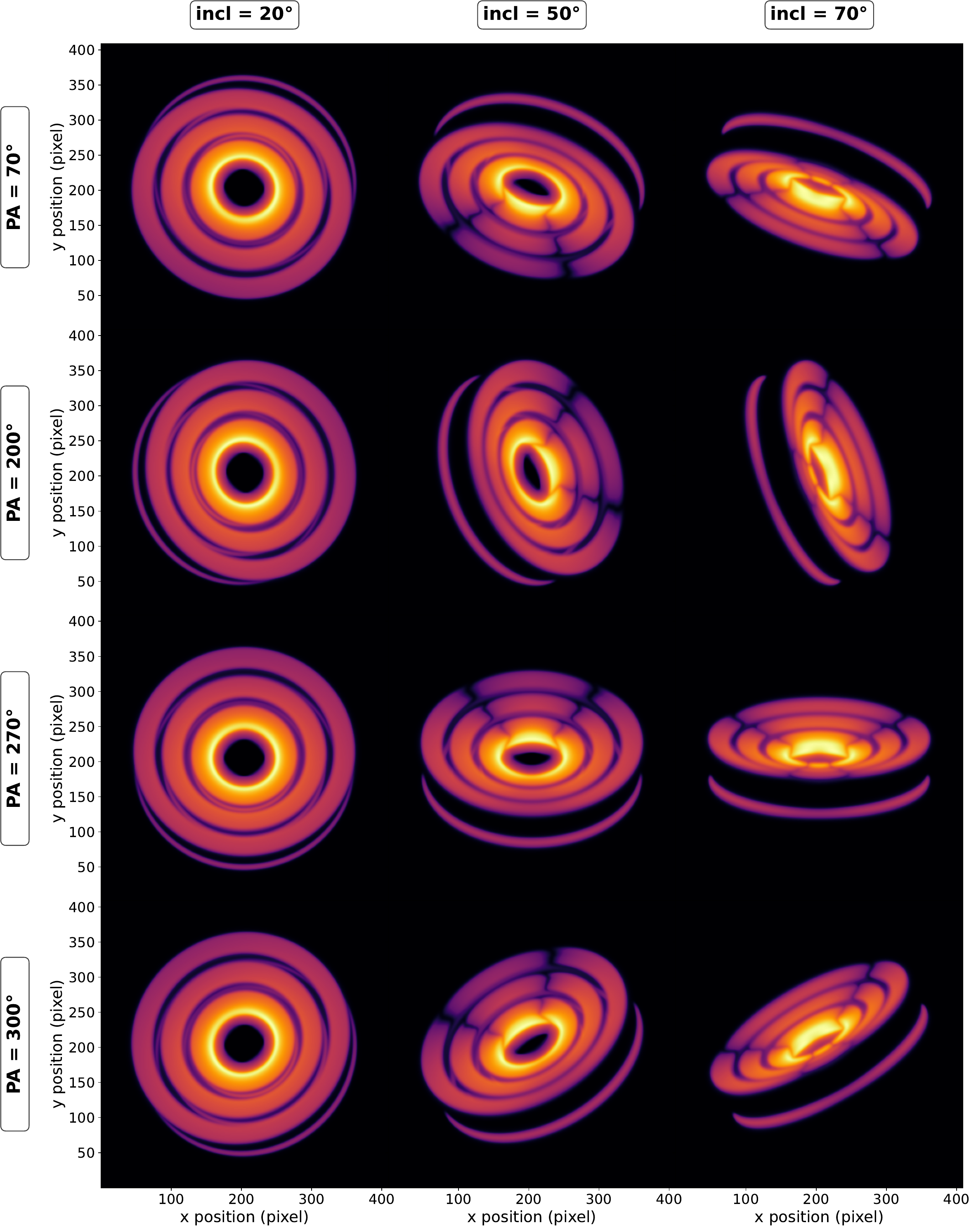}
    \caption{Panel of the twelve simulated disk from MCFOST. All disks have the same structure with 3 rings, but for three inclinations (columns) and four position angle (rows).}
    \label{fig:Models}
\end{figure}

In this section, we display images of the twelve disks simulated with MCFOST (Fig. \ref{fig:Models}) used to test the robustness of our new tool, both for geometric disk estimation and for SPF extraction.
\newpage
\section{Extraction zone impact on SPF extraction} \label{APP:EZimpact}
\begin{figure}[htbp]
    \centering
    \includegraphics[width=\textwidth]{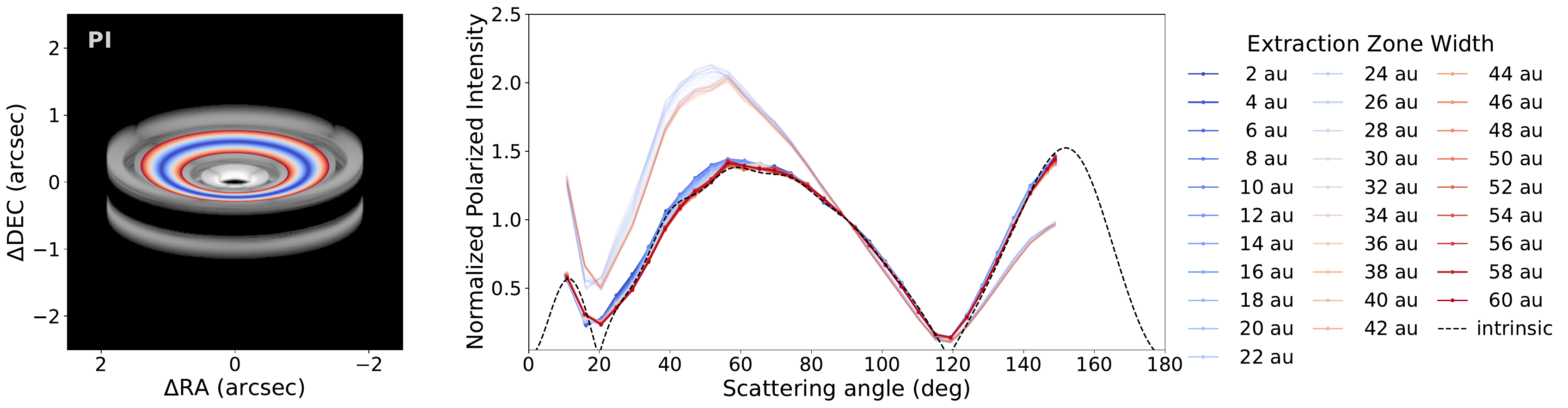}\\
    \includegraphics[width=\textwidth]{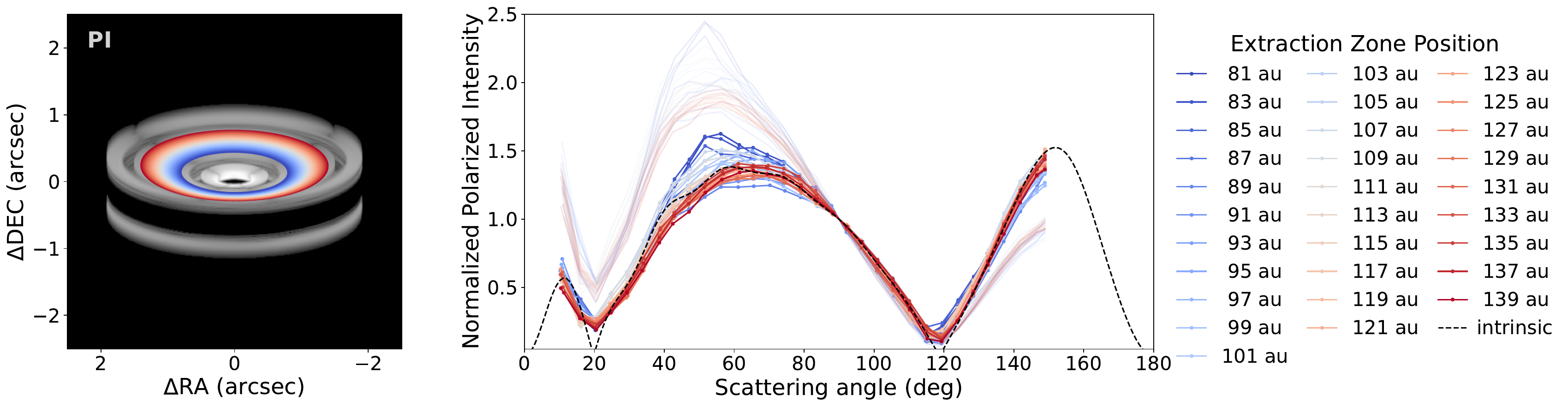}
    \caption{Tests of SPF extraction from a $70$\degr~inclined simulated disk surface with different extraction zones. Top: The extraction zone is centered on the ring and its width increases. Bottom: The extraction zone is $2$ au wide and its radial position increases. Extraction zones are shown as sets of two colored ellipses on the disk image, and corresponding SPF are plotted with the same color. We show the limb brightening corrected SPF in full color, and the uncorrected one in faint color. The theoretical intrinsic dust SPF is displayed in a dotted black line.}
    \label{fig:EZimpact}
\end{figure}

In this section, we investigate the impact of the width and position of the extraction zone on the accuracy of the extracted SPF. The aim is to optimize the positioning of this zone to obtain the most accurate SPF possible. We simulate a protoplanetary disk with three rings at $30$-$50$ au, $70$-$150$ au, and $170$-$190$ au, and we focus on the intermediate one. The disk is inclined at $70$\degr~to cover a wide range of scattering angles, facilitating in-depth comparisons. We first extract the SPF from an extraction zone centered at $110$ au, midway across the ring, and we increase zone width from $2$ to $60$ au. Then we fix the width to $2$ au and move its position from $81$ to $134$ au.\\~\\
First, SPFs extracted for different zone widths are all very similar (see Fig. \ref{fig:EZimpact}, top), showing that the width does not influence the extraction method. However, it should be noted that we are dealing with simulations, and the dust grain properties are the same for the whole ring, which may not be true for a real disk, where a narrower zone will be preferred.\\~\\
Considering now various positions, the SPFs are very similar once again, although there is a noticeable difference when the zone is close to the inner ring edge (see Fig. \ref{fig:EZimpact}, bottom). This suggests that the extraction zone may include areas outside the ring. Surprisingly, this is not the case for the SPF on the outer ring edge. An extraction zone far enough from the ring edges is the optimal choice for good extraction of the disk intrinsic SPF.\\~\\
When extracting the SPF from the disk surface, we assume that surface brightness only depends on scattering angles, neglecting the radial effect linked to the natural brightness decrease proportional to $1/r^2$. In an extended disk, a larger radius leads to a lower surface brightness. However, the resulting error is uniform in each direction and can be globally canceled out, especially for symmetrical disks. 
\newpage

\section{Impact of parameters estimation error on extracted SPF} \label{APP:ParamsImpactSPF}
\begin{figure}[htbp]
    \centering
    \includegraphics[width=\textwidth]{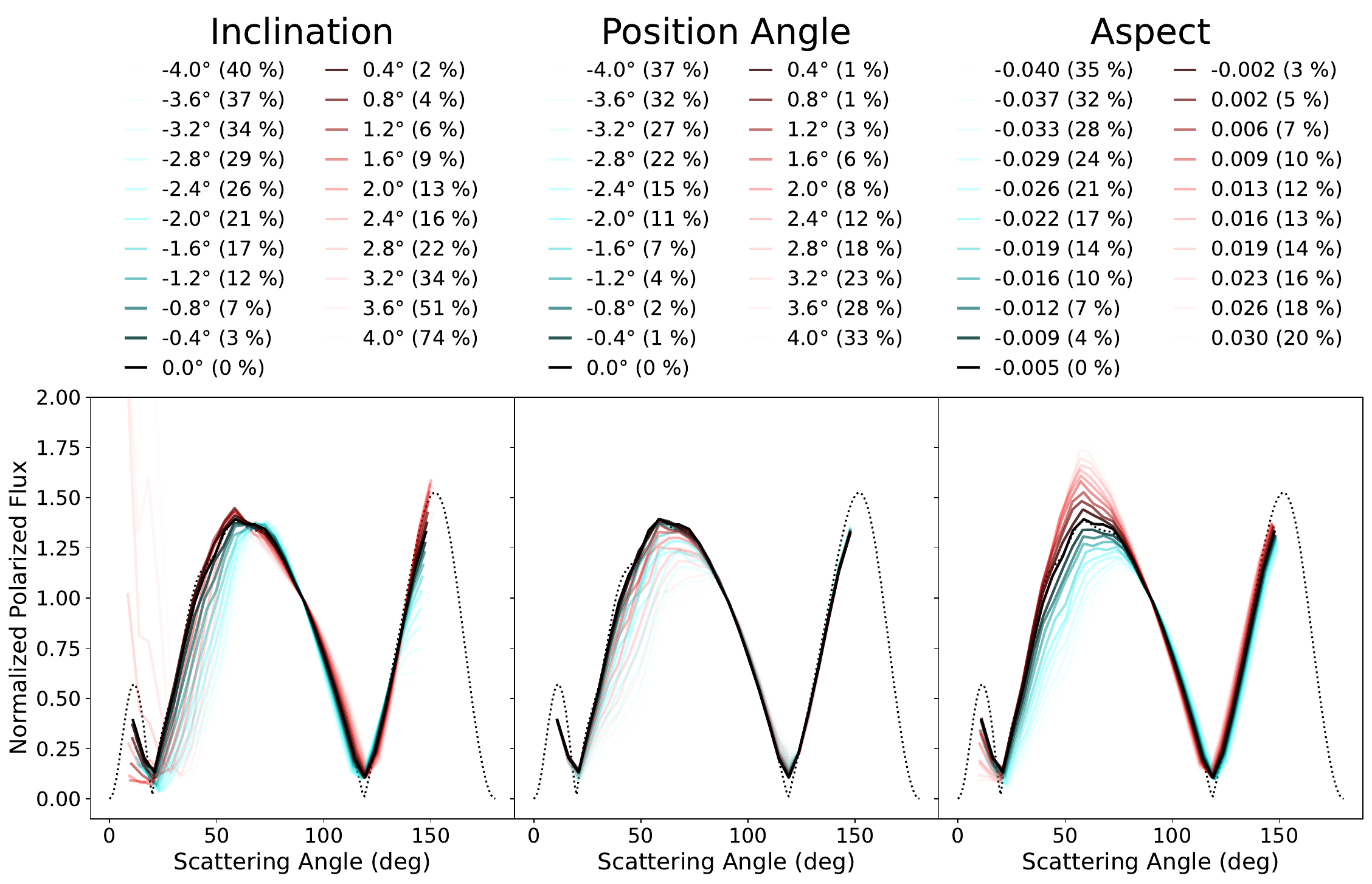}\\
    \caption{Tests of the SPF extraction from a $70$\degr~inclined simulated disk surface with different offsets of $\pm 4$\degr~on inclination (left), $\pm 4$\degr~on position angle (middle), and  $\pm 0.04$ aspect ratio (right). SPFs for underestimated and overestimated parameters are shown in blue, and red, respectively.}
    \label{fig:Error_SPF}
\end{figure}

Errors in the disk geometry parameters affect the accuracy of SPF extraction from the disk surface. In this appendix, we investigate the impact of estimation errors on the extracted SPF, for inclination, position angle, and scattering surface height. Note that the SPFs presented here are all limb brightening corrected, also dependent on geometric estimation. So the study generalizes the impact of estimation error on SPF extraction and correction. We use the same standard model as in section \ref{FittingPart}, inclined at $70$\degr and with a position angle of $270$\degr. The SPFs are extracted by defining extraction zones with different inclinations, position angles, and aspect ratios. We measure the relative error with the reference SPF.\\~\\
All three panels in figure \ref{fig:Error_SPF} show that the SPFs shape is always well recovered, even for errors of $4$\degr~in inclination, $4$\degr~in position angle, and $0.04$ in aspect ratio, especially the $120$\degr~trough.
In particular, an error in inclination results in a shift of the low scattering angle part of SPF towards the higher scattering angles, although the percentage error does not exceed $20\%$ for an error of $\pm 2$\degr~in inclination. The PA primarily alters the amplitude of the SPF peak, and a $2$\degr~error results in an $11\%$ error on the SPF. Finally, the aspect ratio is probably the most sensitive parameter, judging by the SPF extracted. Indeed, the small scattering angle part is strongly impacted by an error on $h/r$, as well as peak amplitude. An error of a few hundredths gives a percentage error of around $15\%$ on the SPF. The largest percentage error is induced by inclination errors, but it should be noted that the computation of the relative error is biased by the shift of the low scattering angle part towards the higher angles. In addition, the shape of the SPF extracted for different inclinations remains very similar to each other.
\newpage
\section{Impact of dust mass on scattering surface height and pressure scale height}\label{APP:DustMass_height}
\begin{figure}[htbp]
    \centering
    \includegraphics[width=\textwidth]{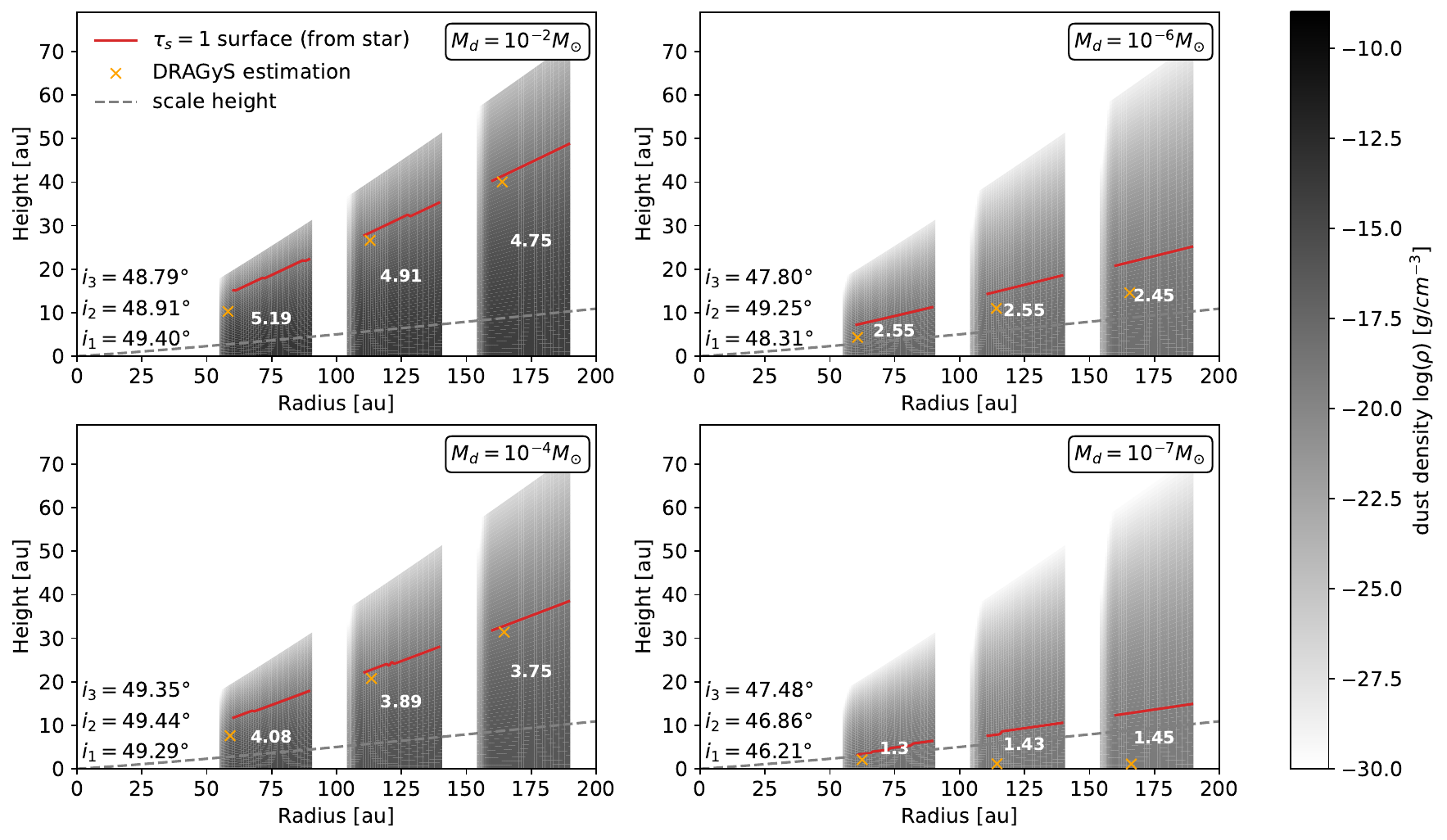}
    \caption{Scattering surface estimation $h_s$ using {\tt{DRAGyS}} (orange cross), compared to the $\tau_s=1$ surface from the star (red curves) for different dust masses from $M_d = 10^{-2}, 10^{-4}, 10^{-6}$ and $ 10^{-7} M_{\odot}$ for each ring (respectively between $60$-$90$ au, $110$-$140$ au, and $160$-$190$ au). As in section \ref{Sect_FitSimu}, we add the dust density map in the background, and we overplot the input gas pressure scale height as a gray dashed curve. The ratio ($\tau_s=1$ surface / pressure scale height) is indicated in white for each ring.}
    \label{APP:Fig_DustMass_height}
\end{figure}

In this section, we investigate the influence of the disk mass on the position of the measured surface with respect to the optical depth $\tau_s=1$. (see Fig. \ref{APP:Fig_DustMass_height}). These tests show that the surface measured by {\tt{DRAGyS}} is very close to the $\tau_s=1$ surface for high disk dust masses ($> 10^{-4}$ M$_\odot$ per ring). We note that the surface estimated by {\tt{DRAGyS}} also moves progressively below the $\tau_s=1$ for the lower disk masses (here $10^{-6}$ and $10^{-7}$ M$_\odot$ in each ring). As expected, the $\tau_s=1$ surface moves closer to the disk midplane for decreasing disk masses and, accordingly, the ratio between the $\tau_s=1$ surface and the pressure scale height also decreases with decreasing disk mass. This ratio also decreases with radius because the pressure scale height increases (power-law index $>$ 1) but the volume density in each ring decreases (because each ring has the same mass but increasing volumes with radius).

\newpage

\section{SPF extraction for different dust properties}\label{APP:SPF_dust_prop}

\begin{figure}[htbp]
    \centering
    \includegraphics[width=\textwidth]{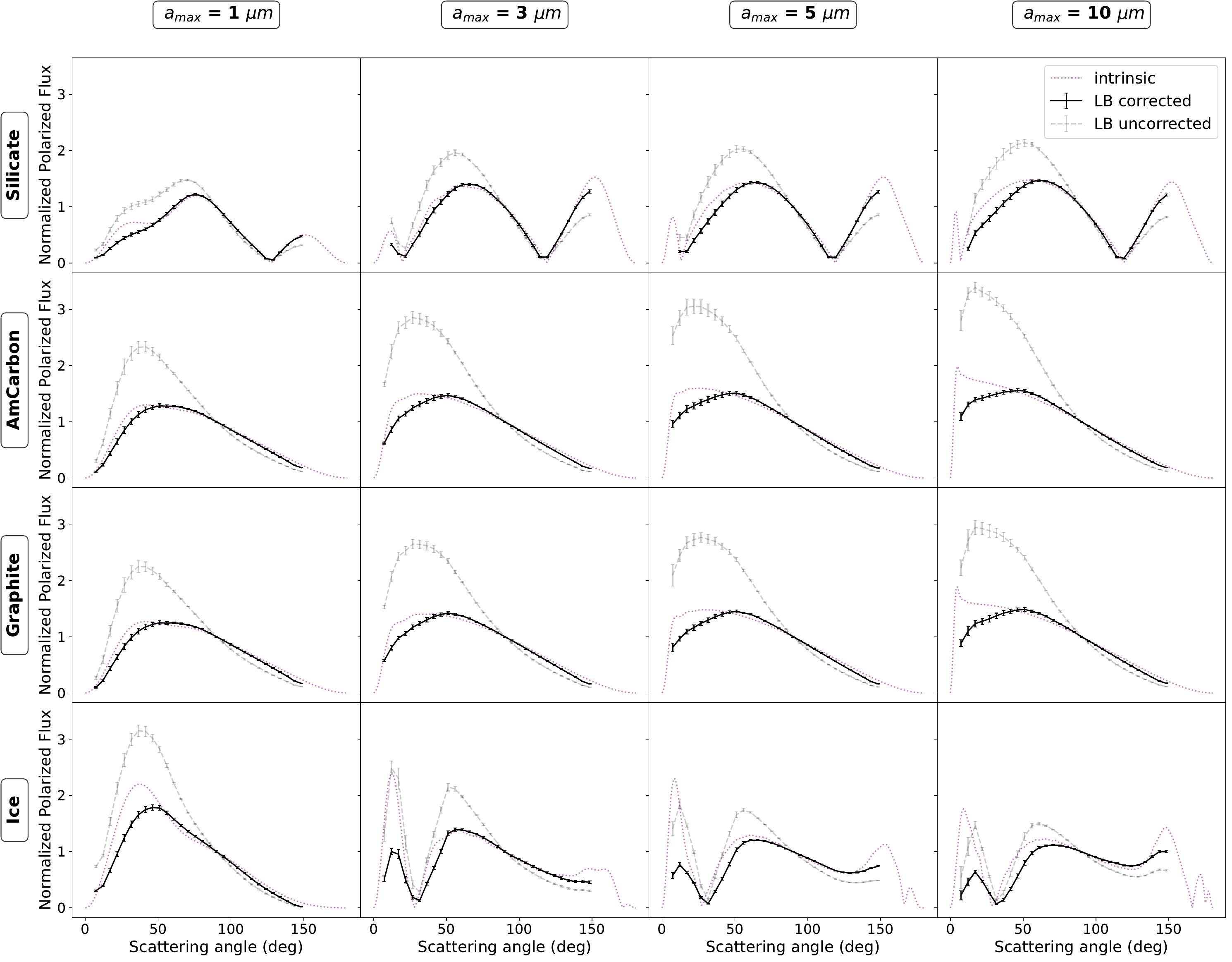}
    \caption{SPFs extracted from simulated disk images using MCFOST, changing the composition (rows) and the size distribution (columns). For each case, we display the intrinsic SPF as the dotted purple curve, and the extracted SPF before and after limb brightening correction as gray and black lines, respectively.}
    \label{APP:Fig_SPF_dust_prop}
\end{figure}

In this section, we investigate the impact of dust composition and grain size distribution on the extracted and limb brightening-corrected SPF. In section \ref{SPFBias}, we showed that applying limb brightening correction yields a remarkable agreement between the extracted SPF and the intrinsic theoretical SPF. However, multiple scattering has the potential to affect the extracted SPF even though this was not the case of silicates with a size distribution ranging from $0.005$ to $3$ $\mu m$ (see figure \ref{SPFSimul}). We therefore performed a study of SPFs extracted from the surface of disks composed of silicate, amorphous carbon, graphite and ice, with maximum grain sizes of $1$, $3$, $5$ and $10$ $\mu m$ (figure \ref{APP:Fig_SPF_dust_prop}).

As we found with our standard model, the correction of the limb brightening effect is a necessity to get closer to the intrinsic SPF. However, it also appears that some SPFs are harder to recover, particularly at small scattering angles, such as for Amorphous carbon, with $a_{max} = 10 \mu m$. This is likely due to the effect of multiple scattering, which tends to depolarize the scattered flux, especially at small scattering angles \citep{tazaki_fractal_2023}.

In conclusion, even if the SPF extracted from our standard model (see Section \ref{S3_SimulationPart}) is in particularly good agreement with intrinsic one, the method we applied generally reproduces well the intrinsic SPFs, although some effects due to multiple scattering can introduce biases at small scattering angles.

\newpage
\section{Detailed SPF for all observations} \label{Apx4_allSPF}
\begin{figure}[htbp]
    \centering
    \includegraphics[width=\textwidth]{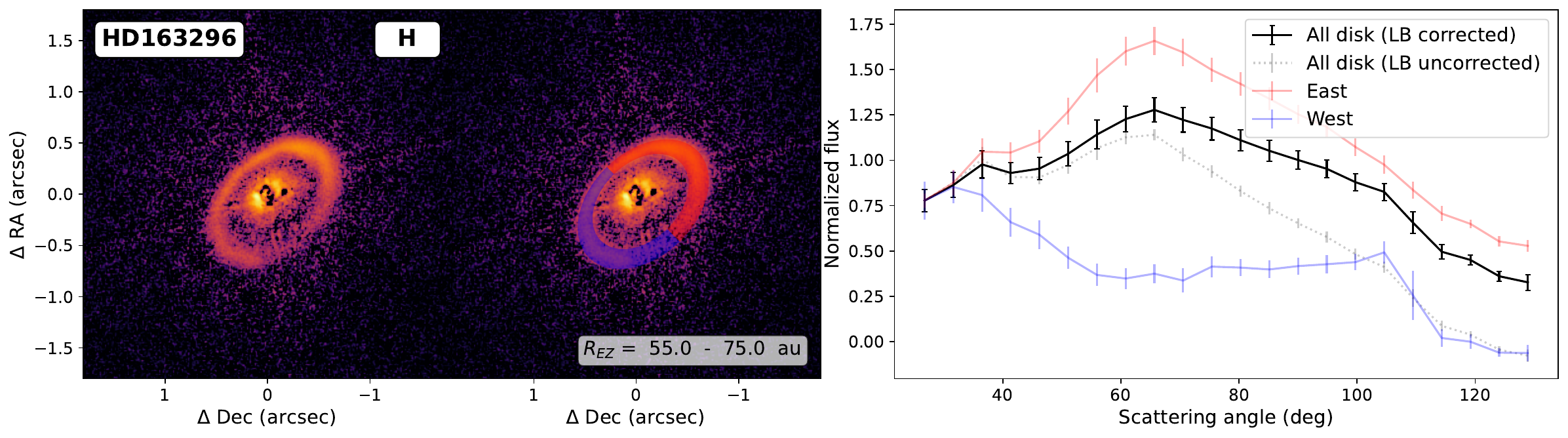}
    \caption{ $Q_{phi}$ image (left) with extraction zone (middle) and extracted polarized SPF (right) for HD 163296 disk in H-Band. We plot SPF for the entire disk (solid black), east side (dotted red), west side (dotted blue), and without correcting for limb brightening effect (dotted grey).}
    \label{APP:SPF_HD163296_H}
\end{figure}

In this section, we present the polarized SPF extracted for each system, including when considering each side of the disk separately. Figures \ref{APP:SPF_HD163296_H}, ... \ref{APP:SPF_V4046_J} correspond to the datasets discussed in section \ref{S4_RealData}.
\begin{figure}[htbp]
    \centering
    \includegraphics[width=\textwidth]{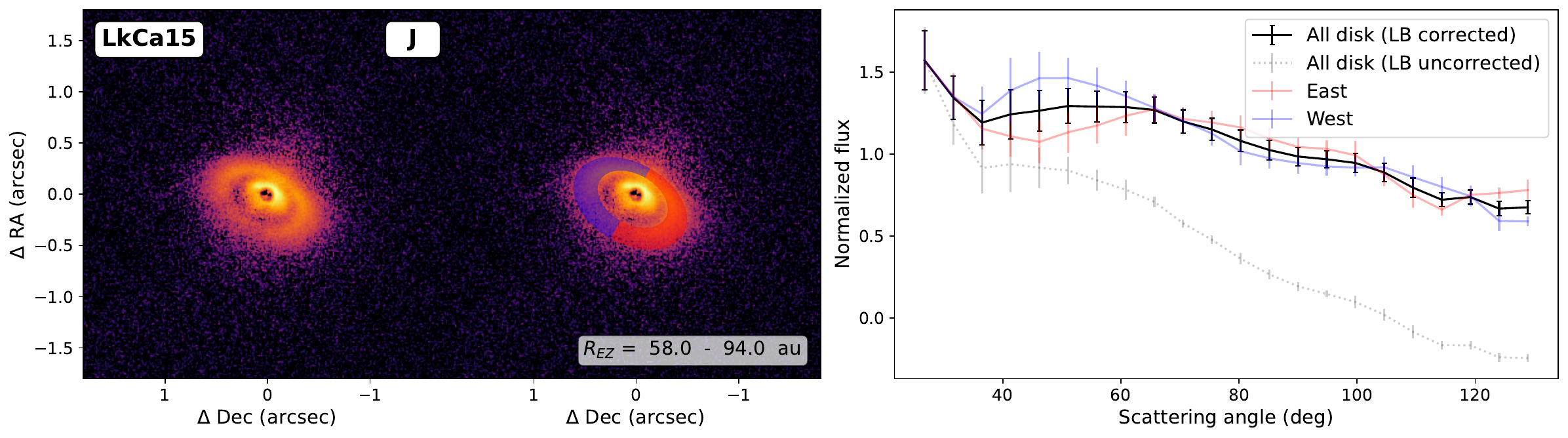}
    \caption{Same as Fig \ref{APP:SPF_HD163296_H} for the LkCa 15 disk in the J band}
    \label{APP:SPF_LkCa_J}
\end{figure}
\begin{figure}[htbp]
    \centering
    \includegraphics[width=\textwidth]{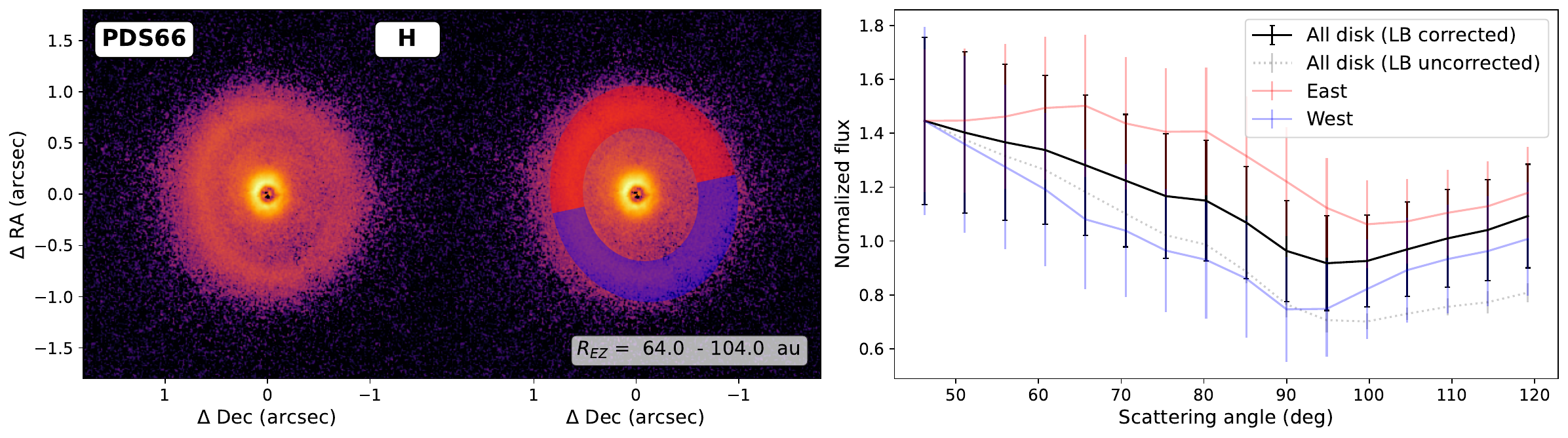}
    \caption{Same as Fig \ref{APP:SPF_HD163296_H} for the PDS 66 disk in the H band}
    \label{APP:SPF_PDS66_H}
\end{figure}
\begin{figure}[htbp]
    \centering
    \includegraphics[width=\textwidth]{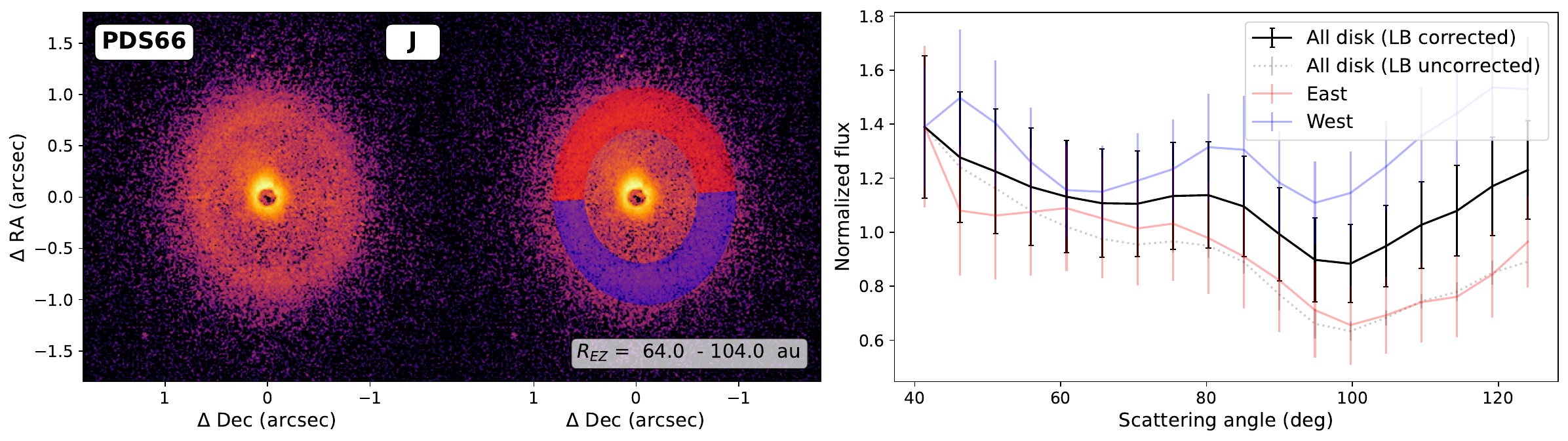}
    \caption{Same as Fig \ref{APP:SPF_HD163296_H} for the PDS 66 disk in the J band}
    \label{APP:SPF_PDS66_J}
\end{figure}
\begin{figure}[htbp]
    \centering
    \includegraphics[width=\textwidth]{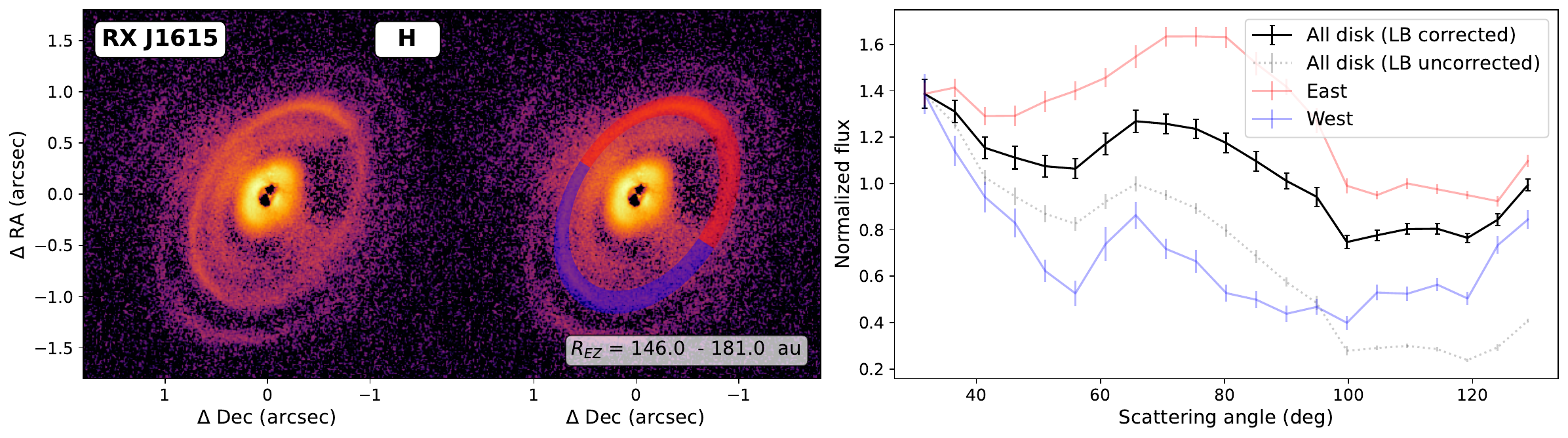}
    \caption{Same as Fig \ref{APP:SPF_HD163296_H} for the RX J1615 disk in the H band}
    \label{APP:SPF_RX J1615_H}
\end{figure}
\begin{figure}[htbp]
    \centering
    \includegraphics[width=\textwidth]{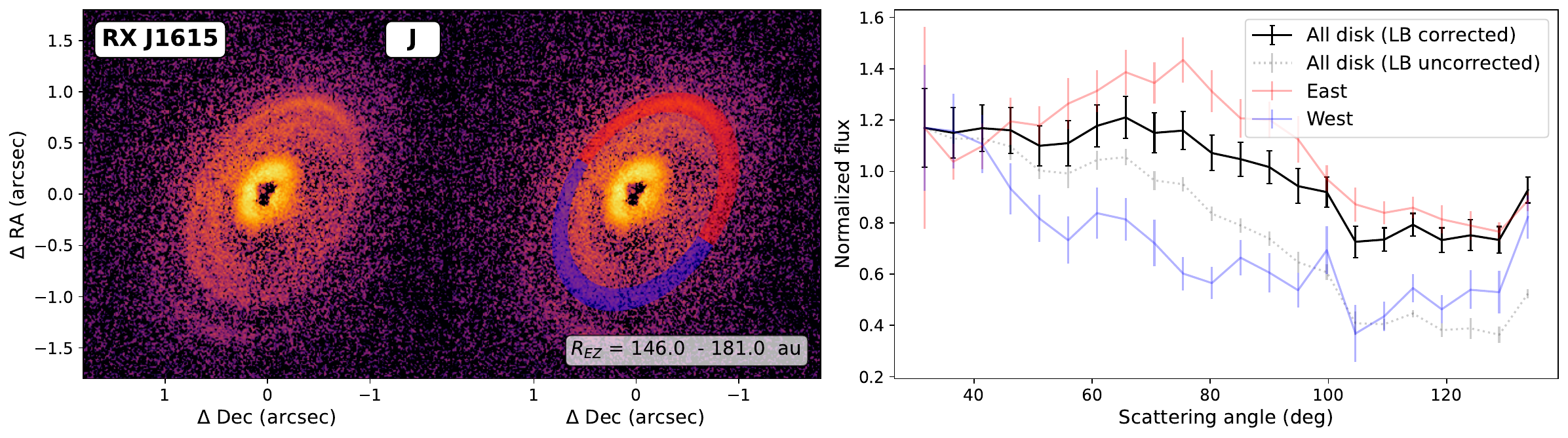}
    \caption{Same as Fig \ref{APP:SPF_HD163296_H} for the RX J1615 disk in the J band}
    \label{APP:SPF_RX J1615_J}
\end{figure}
\begin{figure}[htbp]
    \centering
    \includegraphics[width=\textwidth]{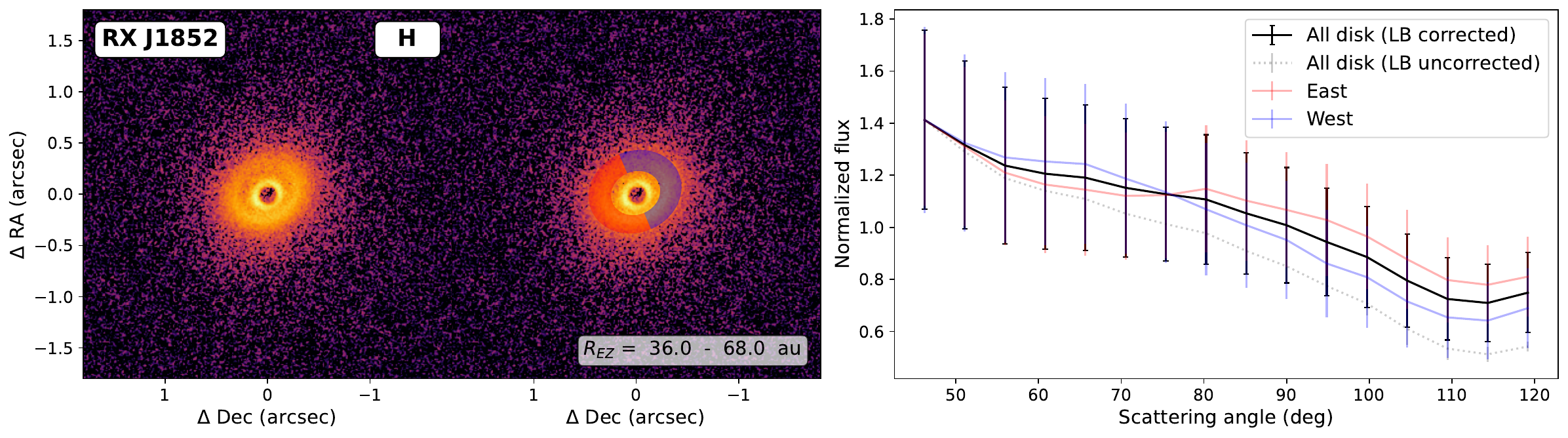}
    \caption{Same as Fig \ref{APP:SPF_HD163296_H} for the RX J1852 disk in the H band}
    \label{APP:SPF_RX J1852_H}
\end{figure}
\begin{figure}[htbp]
    \centering
    \includegraphics[width=\textwidth]{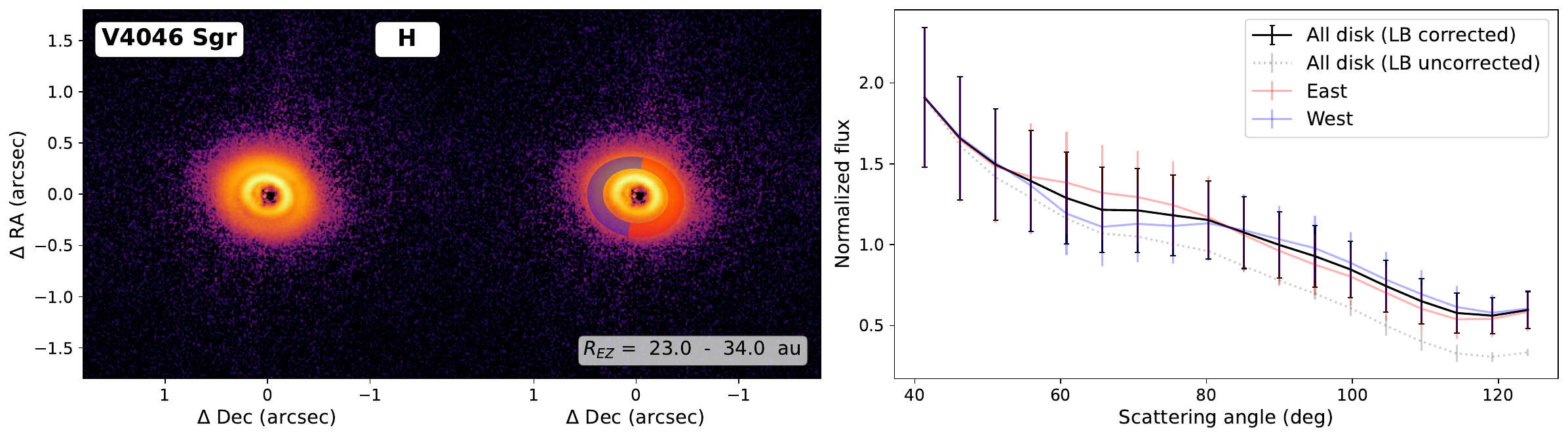}
    \caption{Same as Fig \ref{APP:SPF_HD163296_H} for the V4046 Sgr disk in the H band}
    \label{APP:SPF_V4046_H}
\end{figure}
\begin{figure}[htbp]
    \centering
    \includegraphics[width=\textwidth]{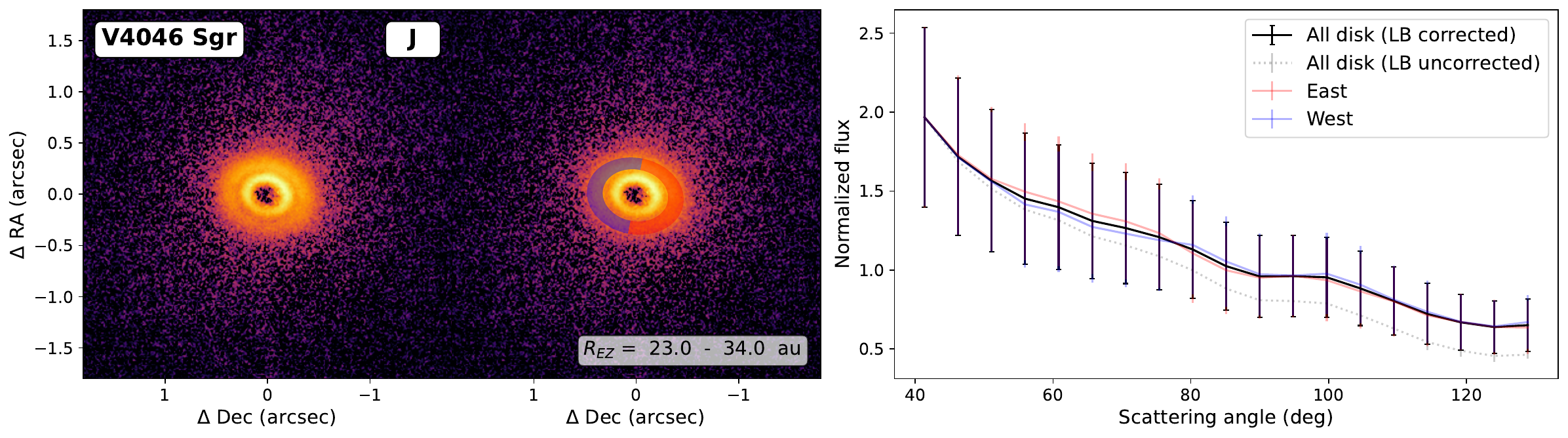}
    \caption{Same as Fig \ref{APP:SPF_HD163296_H} for the V4046 Sgr disk in the J band}
    \label{APP:SPF_V4046_J}
\end{figure}
\end{appendix}

\end{document}